\documentclass[11pt]{JHEP3} 
\usepackage{amsfonts,ulem,graphicx,wrapfig,epsfig}
\newcommand{\be}{\begin{equation}}
\newcommand{\ee}{\end{equation}}
\newcommand{\bea}{\begin{eqnarray}}
\newcommand{\eea}{\end{eqnarray}}
\newcommand{\p}{\partial}
\newcommand{\nn}{\nonumber}
\newcommand{\wt}{\widetilde}
\newcommand{\ra}{\rangle}
\newcommand{\la}{\langle}
\newcommand{\ds}{\displaystyle}
\def\half{\frac{1}{2}}
\def\H{\langle H\rangle}
\def\Hp{\langle H^\perp\rangle}
\def\tH{\langle\wt H\rangle}
\def\tHp{\langle\wt H^\perp\rangle}
\def\bF{\bar F}
\def\bw{\bar w}
\title{
Brane Induced Gravity, its Ghost and   
the Cosmological Constant Problem 
}
\author{S. F. Hassan\\
	Department of Physics \& 
        The Oskar Klein Centre for Cosmoparticle Physics,\\
        Stockholm University, AlbaNova University Centre, 
        SE-106 91 Stockholm, Sweden \\
	E-mail: \email{fawad@fysik.su.se}}
\author{Stefan Hofmann\\
Arnold Sommerfeld Center for Theoretical Physics, LMU,\\ 
Theresienstr. 37, 80333 Munich, Germany\,, \& \\
Excellence Cluster Universe, Boltzmannstr. 2, 85748 Garching, Germany\\
	E-mail: \email{stefan.hofmann@physik.lmu.de}}
\author{Mikael von Strauss\\
	Department of Physics \& 
        The Oskar Klein Centre for Cosmoparticle Physics,\\
        Stockholm University, AlbaNova University Centre, 
        SE-106 91 Stockholm, Sweden \\
	E-mail: \email{mvs@fysik.su.se}}

\abstract{ ``Brane Induced Gravity'' is regarded as a promising
  framework for addressing the cosmological constant problem, but it
  also suffers from a ghost instability for parameter values that make
  it phenomenologically viable. We carry out a detailed analysis of
  codimension $>2$ models employing gauge invariant variables in a
  flat background approximation. It is argued that using instead a
  curved background sourced by the brane would not resolve the ghost
  issue, unless a very specific condition is satisfied (if satisfiable
  at all). As for other properties of the model, from an explicit
  analysis of the 4-dimensional graviton propagator we extract a mass,
  a decay width and a momentum dependent modification of the
  gravitational coupling for the spin 2 mode. In the flat space
  approximation, the mass of the problematic spin 0 ghost is
  instrumental in filtering out a brane cosmological constant. The
  mass replaces a background curvature that would have had the same
  function. The optical theorem is used to demonstrate the suppression
  of graviton leakage into the uncompactified bulk. Then, we derive
  the $4$-dimensional effective action for gravity and show that
  general covariance is spontaneously broken by the bulk-brane setup.
  This provides a natural realization of the gravitational Higgs
  mechanism. We also show that the addition of extrinsic curvature
  dependent terms has no bearing on linearized brane gravity.}


\preprint{}

\begin{document} 

\section{Introduction and Results}

``Brane induced gravity'' (BIG)\cite{DGP,DG} (for a review see
\cite{G-rev}) is regarded as providing one of the few promising
frameworks for addressing the cosmological constant problem
\cite{Weinberg} in the sense of explaining how the observed value of
vacuum energy could be so small as compared to the types of values
inferred from quantum field theory \cite{DGS1,DGS2} (other approaches
include \cite{Bousso,Demir}, for reviews see
\cite{Polchinski-rev,Nobbenhuis1,Nobbenhuis2}). In its basic form, the
model regards the $4$-dimensional universe as a $3$-brane in a
non-compact $(4+n)$-dimensional bulk spacetime. Gravitational dynamics
arise from Einstein-Hilbert actions both in the bulk and on the brane.
Bulk gravity absorbs the effect of brane cosmological constant,
leaving behind effectively $4$-dimensional gravity on the brane with a
``filtered out'' cosmological constant. Unfortunately consistent
constructions are still lacking, for example, issues related to the
presence of a tachyonic ghost mode have not been resolved
satisfactorily. Hence more work is needed to obtain {\it consistent}
working models of this type\footnote{In this paper we will not discuss
  variants of the basic setup like cascading gravity
  \cite{deRham:2007rw} that address the ghost issue
  \cite{deRham:2010rw}.}.

\subsection{An overview of BIG, its promise and problems}

Let's briefly review the origin, workings and shortcomings of the
model. 

{\bf Historical background and motivation:} In ordinary
Einstein-Hilbert gravity a cosmological constant $\Lambda$ in a $3+1$
dimensional spacetime gives rise to a de Sitter metric. However, if
this specetime is a $3$-brane embedded in a $4+n$ dimensional bulk,
then it is known that bulk gravity has classical solutions in which
$\Lambda$ (which is the same as the brane tension) curves directions
transverse to the brane while the metric along the $3$-brane remains
flat, provided $n>1$ \cite{Gregory:1995qh,Charmousis, Rub-Shap}. This
eliminates $\Lambda$ from the brane point of view at the level of
classical solutions. But in this model, gravity is a
$(4+n)$-dimensional force. 

To get effectively $4$-dimensional gravity on the brane without
compactifying the extra dimensions, Dvali, Gabadadze and Porrati
\cite{DGP,DG} proposed adding to the bulk gravity action, a
$4$-dimensional Einstein-Hilbert action for the brane induced metric.
Such brane localized actions, along with extra corrections, naturally
arise in field theory \cite{DGHS} as well as in string theory
\cite{Corley,Ardalan,Antoniadis,Kohlprath,Epple, Kiritsis} setups,
hence the name {\it brane induced gravity}. However, for
phenomenological reasons, the parameters have to be tuned to make the
brane localized action dominant. The $n=1$ case is the well studied
DGP model \cite{DGP,G-rev}. It is special in that it does not suffer
from tachyon/ghost problems. But it is also not adequate for
addressing the cosmological constant problem in the sense stated
above. In \cite{DGS1,DGS2} it was argued that models with $n>2$ should
be able to filter out a brane cosmological constant, making gravity on
the brane less sensitive to it, while at the same time giving rise to
essentially $4$ dimensional gravitational interactions on the
brane. These arguments were mainly based on the classical solutions of
\cite{Gregory:1995qh,Charmousis} described above \footnote{The $n=2$
  models also filter out $\Lambda$ but they share features of both
  $n=1$ and $n >2$ models and are discussed in some detail in
  \cite{Corradini,Kaloper1,Kaloper2}.}.

Beside the classical solutions, most of the explicit work on BIG with
$n>2$ has focused, for reasons of technical simplicity, on the
$\Lambda=0$ case, analyzing the propagation of linearized metric
fluctuations for a flat brane, in flat bulk spacetime. The purpose is
to, from this, glean information on infrared modifications of gravity
that would eventually result in filtering out $\Lambda$. In
particular, one finds that for appropriate values of parameters in the
action, gravity can be made effectively $4$-dimensional over a desired
range of distances. It will also contain massive unstable gravitons
\cite{DG,DGS1,DGS2,GS}. Beyond this range (say, for distances of the
order of the size of the Universe), the emergent unstable graviton,
with a small but momentum dependent mass, produces infrared
modifications of gravity. This is supposed to be relevant to filtering
out $\Lambda$, as suggested by the classical solutions mentioned
above. 

{\bf Problematic issues:} The above picture looks promising,
except for some important issues that arise in the analysis carried
out so far, and that need to be better understood and resolved. We
describe five issues below. 

\begin{enumerate}
\item The evidence in \cite{DGS1,DGS2} that BIG would filter out a brane
cosmological constant is mostly based on the existence of the
classical solutions of \cite{Gregory:1995qh,Charmousis} that are
supposed to describe how $\Lambda$ curves the space transverse to the
brane. However a closer examination shows that in the vicinity of the
brane, the source structure implied by these solutions does not
correspond to the brane source (which is a common problem of brane
solutions). This means that a detailed analysis of fluctuations
(needed to obtain the effective $4$-dimensional gravity on the brane)
cannot be carried out around existing classical
solutions \footnote{Progress in this direction will be reported
  elsewhere.}. To avoid this problem, in this paper we start with a
flat background approximation and try to glean information about the
effects of background curvature sourced by the brane.

\item Second, as pointed out in \cite{DR}, the analysis of BIG in a flat
background approximation indicates the presence of a tachyonic ghost
mode and, therefore, an inconsistency of the model. This is a major
drawback and resolving the ghost problem is crucial for developing
these models further. It is easy to see that a ghost generically
appears in massive gravity theories unless it is given an infinite
mass, as in Fierz-Pauli gravity \cite{FP1,FP2,vDVZ1,vDVZ2,vDVZ3}. But
in BIG around flat background, the {\it tachyonic ghost} cannot be
avoided easily. There have been attempts to evade the problem by
modifying the setup \cite{deRham:2010rw,GS,Porrati}, but a
satisfactory resolution is still lacking. 

\item The third issue is the mode of realization of the gauge symmetry,
in this case $4$-dimensional general covariance on the brane. This is
related to the structure of the effective theory on the brane. It is
known that in both Fierz-Pauli massive gravity, as well as in Higgs
gravity setups (see \cite{'tHooft, Kakushadze}), a graviton mass is
related to the breaking of $4$-d general covariance. However, in BIG,
the starting theory is manifestly invariant under bulk and brane gauge
transformations, while at the same time $4$-d gravitons acquire an
effective mass. In usual treatments this result is obtained after gauge
fixing which obscures the nature of symmetries of the effective theory. 
How is the generation of mass related to the realization of symmetry
in BIG models?

\item The fourth issue is a more detailed understanding of the technical
aspects of the resulting modified gravity on the brane and its
response to the cosmological constant, both in flat and curved
background approximations. Technically, a zero thickness brane should
consistently be replaced by one with an effective width
\cite{DG,DGHS,Kiritsis}, before a quantitative investigation of
graviton mass, its decay and other possible modifications of the
gravitational force. It is also important to see explicitly how a
cosmological constant affects the situation. In this paper such an
analysis is carried out in the flat background approximation.

\item Related to this is the fifth issue of naturalness of scales in
BIG, at least based on some elementary considerations. The basic BIG
model contains bulk and brane Planck scales, say, $M_*$ and $M_P$,
respectively, as free parameters. For a realistic $M_P\sim
10^{19}GeV$, to insure that the resulting IR modifications of gravity
are not too large, the bulk gravity scale $M_*$ should be unnaturally
low ($<10^{-3}eV$ for $n\geq 2$), corresponding to strongly coupled
bulk gravity \cite{DG}. The effective brane thickness introduces
another scale, though often this is taken to be related to $M_*$
\cite{DGHS, Kiritsis}. But a non-zero $\Lambda$ also affects the
identification of the gravitational constant and other standard model
couplings, influencing the problem of scales. Taking this into account
is important since as a viable theory BIG should remain calculable at
the interacting level and should not generate new hierarchy problems. 
It is not obvious that BIG could meet these requirements.
\end{enumerate}

Considering the potential ramifications of brane induced gravity for
the cosmological constant problem, we revisit these models and analyze
them in some depth in the hope of gaining a better understanding of
the issues outlined above. This work address some of these issues
directly, and hopefully sheds light on, and sharpens the context of,
the rest. 

\subsection{Overview and discussion of our results}

This subsection contains a description of our results and their
implications. We concentrate mainly on codimension $n>2$ BIG models
relevant to the cosmological constant problem and allow for the
addition of extrinsic curvature terms and a brane cosmological
constant (brane tension) $\Lambda$. In this model, we study linearized
metric fluctuations around a background configuration which is
approximated as a flat $3$-brane in a flat bulk spacetime (for the
reason explained under item (1) of the previous subsection). The
crucial point is that the presence of a small brane cosmological
constant does not invalidate the flat background approximation.

{\bf Spontaneous breaking of general covariance:} The basic variables
are the bulk metric $G_{MN}$ and the brane embedding functions
$x^M(\sigma)$. The symmetries are the brane and bulk general
coordinate transformations, involving $\sigma^\mu$ and $X^M$,
respectively. The standard gauge choice (Monge or static gauge for the
brane, and harmonic gauge for the bulk symmetries) is natural for bulk
physics but obscures the nature of symmetries of the effective brane
theory. Thus, with a focus on $4$-dimensional physics, we work with
gauge invariant variables and the induced metric
$g_{\mu\nu}=G_{MN}\p_\mu x^M\p_\nu x^N$, that naturally couples to
brane matter. The outcome is that, beside the usual gauge invariant
parts of the metric, we encounter $4$ new gauge invariant variables.
These are closely related to St\"uckelberg fields used to restore
gauge invariance in Fierz-Pauli massive gravity. The $4$-dimensional
effective action, obtained after integrating out all extra dimensional
modes, depends on these variables, and through them, on gauge
dependent components of the metric. This explicitly demonstrates how
$4$-dimensional general covariance is spontaneously broken by the
bulk-brane setup in the effective $4$-dimensional theory, whereas all
symmetries are manifest in the starting action. Thus, BIG naturally
implement a gravitational Higgs mechanism of the sort discussed in
\cite{'tHooft,Kakushadze}. These issues are discussed in sections $6$
and $7$.

{\bf Explicit analysis of graviton mass and decay:} To study graviton 
propagation on the brane, one has to consider thick branes to avoid
divergences associated with vanishing brane width.  It is then known
that integrating out bulk modes contributes a ``self-energy'' term to
the $4$-dimensional graviton propagator, giving, schematically (see
equations (\ref{s},\ref{hp},\ref{GSigma})),
\be
\frac{O_{\mu\nu\rho\sigma}^{(0,2)}}{Bk^2+A\omega^{n-2}N^{(0,2)}
  \Sigma(k)}\,.    
\label{schematic}
\ee 
$A$ and $B$ are inverse Newton constants in $d$ and $4$ dimensions,
$\omega$ is the brane width and the superscripts refer to spin $0$ and
$2$ modes of graviton. The scale of graviton mass is set by
$A\omega^{n-2}/B$, taken to be very small for phenomenological
reasons. 

\begin{enumerate}
\item After reviewing the well known problem of the tachyonic ghost
nature of the spin $0$ mode, we use a Gaussian profile for the brane
thickness to explicitly extract from $\Sigma(k)$ a mass term, a decay
width and a momentum dependent modification of the Newton constant for
the healthy spin $2$ mode. All this hinges on the fact that, as we
show, the standard perturbative treatment of unstable particles in
quantum field theory remains applicable for small graviton masses.

\item The suppression of the probability for graviton escape into the
bulk, resulting in effectively $4$-dimensional gravity on the brane,
is demonstrated using the optical theorem. 
\end{enumerate}

{\bf Suppression of cosmological constant:} In Einstein-Hilbert
gravity, a cosmological constant $\Lambda$, however small,
destabilizes flat space, giving rise to de Sitter or anti de Sitter
spacetimes. In BIG, a brane $\Lambda$ can curve directions transverse
to the brane, leaving the brane metric flat. However, in the analysis
of BIG around a {\it flat background}, commonly employed to extract
more detailed information about brane gravity, the setup does not
allow for absorbing $\Lambda$ in the curvature of extra dimensions.
But, as discussed in section 4, the theory compensates for this and
neutralizes $\Lambda$ using a $\Sigma(0)\neq 0$ piece of the
self-energy in the scalar mode propagator, closely related to its
tachyonic mass. This leaves the brane metric flat\footnote{The $n>2$
  BIG only attempts to explain why one does not observe a large
  cosmological constant. At least in this form, it does not explain
  the observed recent accelerated expansion due to a very small
  $\Lambda$.}, and only modifies gravitational and non-gravitational
couplings. Based on this analysis, the following observations should
be made: 

\begin{enumerate}
\item As far as filtering out $\Lambda$ is concerned, the tachyonic mass
of the scalar ghost replaces the function of a background curvature
(as an artifact of flat background approximation) and is not
unhealthy. In fact, the mass is tachyonic only with respect to the
wrong sign of the ghost kinetic term, but has a positive contribution
to the energy.

\item The momentum dependence of $\Sigma(k)$, which determines the
graviton decay width and normalizes the gravitational coupling, has no
bearing on the suppression of $\Lambda$ (which is entirely due to
$\Sigma(0)$). This is to be contrasted with the more general {\it
  degravitation} mechanisms \cite{ADDG} where a momentum dependent
gravitational coupling could make the theory less sensitive to
$\Lambda$. But the outcome of $n>2$ BIG is closer to the scenario in
\cite{DHK}, more closely modeled around massive gravity.

\item In some of the literature on BIG, the suppression of $\Lambda$ is
heuristically explained in terms of $3$-brane gravity becoming
$4+n$-dimensional, hence weaker, at very large distances. This truly
is the behaviour of $n=1$ theory which, however, does not filter out
$\Lambda$. A closer inspection of the propagator in $n>2$ models shows
that brane gravity at large distances remains $4$-dimensional and is
dominated by $\Sigma(0)$, related to graviton mass, which suppresses
$\Lambda$.

\item A lesson from this analysis is that the flat background
approximation to BIG remains valid in the presence of a small
cosmological constant, with implications for the resolution of the
ghost problem to be discussed below.
\end{enumerate}

{\bf Extrinsic curvature effects:} Derivations of the BIG action in 
string theory setups also produce terms dependent on the extrinsic
curvature of the brane. One may wonder if neglecting these terms has
an effect on the  ghost problem. The analysis in section $5$ shows
that these terms have no effect on brane gravity at the linearized
level. This may be disappointing from the point of view of the ghost
problem but also shows that these terms do not drastically modify
gravity on the brane. 

{\bf Status of the ghost problem:} In equation (\ref{schematic}) for
the graviton propagators in BIG, $O_{\mu\nu\rho\sigma}^{(0,2)}$
contain the tensor structure. This is exactly the same as the tensor
structure of Einstein-Hilbert gravity in $4$ dimensions (obtainable by
setting $A=0$). It is known that with this tensor structure, the
theory is ghost free only for zero masses. Any non-zero mass pole
implies a ghost, as is the case with BIG\footnote{The ghost can also
  be avoided by changing the tensor structure gravity in BIG for
  example, as in \cite{GS,Porrati} but that introduces vDVZ type
  discontinuity in the free theory. These could be avoided in the
  non-linear regime by the Vainshtein mechanism
  \cite{Vainshtein,Deffayet}, but we will not consider such
  alternatives here.}. Resolving this problem is crucial for any
eventual application of BIG. We do not achieve this in this paper, but
sharpen the context of the problem.

\begin{enumerate}
\item As a warm up consider the following puzzle: The BIG action can
arise within seemingly consistent setups, like the string theory setup
of \cite{Corley}. How can these consistent setups lead to a ghost
inconsistency? To answer this, note that the identification of ghost
in BIG hinges on the $\Sigma$ term in (\ref{schematic}) being a {\it
  small} self-energy correction. This is obviously the case in the
phenomenologically interesting regime of small graviton mass ($\sim
A\omega^{n-2}/B$) which requires that the brane Einstein-Hilbert
action dominates over the bulk one. If this is not the case, then
either the ghost may have a large mass, beyond the regime of validity
of the low-energy effective action, or $\Sigma(k)$ may contain
appreciable corrections to $Bk^2$ making the above interpretation of
the propagator meaningless. This is the case in the string theory
regime of \cite{Corley} where, for a small string coupling $g_s$,
$A\sim 1/g_s^2$ dominates over $B\sim 1/g_s$ (assuming $\omega\sim
g_s^0$), the propagator is essentially $\Sigma^{-1}(k)$, and gravity
is $d$-dimensional and ghost free. On the other hand, assuming
$\omega^4\sim g_s$ and $n=4$ (more natural for D3-branes), the mass is
at string scale.

\item Could the appearance of ghost (in the phenomenologically
interesting cases) be related to the use of flat background
approximation, ignoring $\Lambda$ and its back reaction on bulk
geometry? As discussed in section $4$, since our analysis is already
valid for a small $\Lambda$, one can answer this in the negative by a
continuity argument: that in any case, the small $\Lambda$ limit
should correspond to our result which is not ghost free.

One may formulate a sharper criterion: In a gravitational
background sourced by a brane $\Lambda$, the metric 
along the brane remains flat, hence the background mostly affects
gravitational dynamics in the bulk. In terms of the propagators in
(\ref{schematic}), the curved background mostly affects the
$A$-dependent self-energy term $\Sigma$, leaving $B k^2$ and tensor
structure unchanged. Then in the phenomenologically interesting regime
of small mass, where the $B$ term dominates, the propagator has the
same structure as in flat background. The discussion above then shows
that the ghost mode can be avoided only if the new self-energy term
satisfies $\Sigma_{curved}(0)=0$, implying no hard masses. This can be
regarded as the constraint on admissible ghost free backgrounds, if
any. For more on this see the concluding section.

\item In the $4$-dimensional effective action for BIG derived in section
$7$, the ghost can be related to the St\"uckelberg-like fields that
arise in the gauge invariant formalism, and hence to the spontaneous
breakdown of general covariance. This underlines the similarity with
the ghost in standard massive gravity and in Higgs gravity.
\end{enumerate}

\subsection{Organization of the paper}

The rest of the paper is organized as follows: In section 2, we
introduce the model and detail the setup for the analysis. Gauge
invariant variables are introduced and a brane thickness profile is
used to avoid divergences associated with zero-thickness branes.

In section 3, we solve the equations for the basic BIG model and
review the tachyonic ghost problem. The suppression of graviton escape
into bulk is described in terms of the optical theorem. We then carry
out an explicit analysis of the graviton propagator, extracting from
it a mass, a decay width and a modulation of the gravitational
coupling. It is shown in the process that standard perturbative QFT
methods remain applicable for small graviton masses.

In section 4, we consider the effect of a small brane cosmological
constant $\Lambda$ in the flat background approximation, showing that
for codimension $n>2$, it is filtered out by the tachyonic scalar
mass. The tachyon problem is dismissed. The ghost problem is argued to
survive, even when the theory is analyzed in a background sourced by
$\Lambda$, except for special backgrounds (if they exist at all) that
do not make brane gravity massive. 

In section 5, we solve the equations in the presence of extrinsic
curvature terms that generically arise in derivations of the BIG
action. It is shown that they have no effect whatsoever on the brane
gravity. 

In section 6, we revisit massive Fierz-Pauli gravity with
St\"uckelberg fields in terms of gauge invariant variables. In
particular, it is shown that the correct solutions can be obtained
from a $1$-parameter family of actions. 

In section 7, we derive an effective action for the brane gravity by
integrating out bulk related modes. We construct a $1$-parameter 
family of effective actions that reproduce the correct solution and,
generically, contain St\"uckelberg fields. This shows that gauge
dependent modes of metric do not decouple, implying a spontaneous
breakdown of $4$-dimensional general covariance. The relation to
the gravitational Higgs mechanism is also pointed out.
The conclusions are summarized in section 8, and the appendix contains
a brief discussion of gauge invariant variables and related zero mode
issues.   

\section{Preliminaries}
In this section we introduce the {\it brane induced gravity} model and 
describe our setup for its analysis: the use of gauge invariant
variables, description of thick branes in terms of a density profile
and its use as a regulator. Finally we consider the bulk-brane
relations and the appearance of $4$ new gauge invariant variables
peculiar to the bulk-brane setup.   

\subsection{The model and its origins}

In brane induced gravity models, the Universe is regarded as a
$3$-brane with coordinates $\sigma^\mu$ ($\mu=0,\cdots,3$) embedded in
a non-compact $(d=4+n)$-dimensional bulk spacetime with coordinates
$x^M$ through $x^M(\sigma)$. In the {\it basic} model, dynamics of
gravity is governed by an Einstein-Hilbert action $S^{bulk}_{EH}$ for
the bulk metric $G_{MN}(x)$ and a similar action $S^{brane}_{EH}$ for
the brane induced metric, $g_{\mu\nu}=\p_\mu x^M\p_\nu x^N G_{MN}$,
sourced by brane matter $S^{brane}_m$ \cite{DGP,DG}.  In this paper,
we also allow for adding a brane tension (brane cosmological constant)
term $S^{brane}_\Lambda$ and extrinsic curvature corrections
$S^{brane}_\Omega$, to be specified later. Then\footnote{We follow the
  sign conventions of Weinberg in \cite{Weinberg-book}},
\be 
S=-A\int d^dx\sqrt{-G}R^{(d)}-B\int d^4\sigma\sqrt{-g}R^{(4)}+
S^{brane}_\Omega+ S^{brane}_\Lambda + S^{brane}_m \,.
\label{Sbig-I}
\ee 
As argued in \cite{DGP,DG,DGHS}, such models naturally arise in
braneworld setups as a result of integrating out massive matter on the
brane following \cite{Adler}. More specifically, they were shown to
arise in non-supersymmetric string theory, as a result of closed
string scattering off D-branes \cite{Corley,Ardalan} where the action
(\ref{Sbig-I}), including $S^{brane}_\Omega$ can be computed
explicitly\footnote{The calculation of \cite{Corley,Ardalan} leads to
  the expected behaviour of $A\sim g_s^{-2}$ and $B\sim g_s^{-1}$ with
  the string coupling $g_s$. Mass dimensions are supplied by the
  string tension, $1/\alpha'$. A curious feature of their result is
  that while, $A>0$, one gets $B<0$. Also, the relative sign between
  $S^{brane}_{EH}$ and $S^{brane}_\Omega$ is opposite to what appears
  in the Gauss-Codazzi equation. However, for the BIG results to be
  relevant to phenomenology, we must take $B>0$ and adjust the
  parameter values appropriately, as will be specified later.}. At
this order in the $\alpha'$ perturbation theory, there are also bulk
$R^2$ terms, in the Gauss-Bonnet combination, that do not contribute
to our analysis, as well as terms involving other massless string
states that are ignored here. A different string theory realization
was proposed in \cite{Antoniadis}. Here for the  couplings $A$, $B$,
{\it etc.} we do not use their calculated high energy values, but
treat them as low energy phenomenological parameters. 

The $d=5$ model is the well known DGP model that has been extensively
investigated.  As argued in \cite{DGS1,DGS2} it cannot address the
problem of the observed smallness of cosmological constant.  $d=6$ is
also special and was investigated in
\cite{Corradini,Kaloper1,Kaloper2}. Below, we are interested mainly in
$d>6$ models.

\subsection{Flat background and gauge invariant variables}

We investigate brane induced gravity at the linearized level around a
flat background, which is technically the easiest and a good first
approximation. Consider a flat bulk with a flat brane and split
$\{x^M\}$ into $\{x^\mu_{||}\,,x^i_{\perp}\}$, respectively, parallel
and perpendicular to the brane. The flat brane corresponds to 
$x^\mu_{||}=\sigma^\mu$, $x^i_{\perp}=y^i_0$ (constants). Linearizing
$G_{MN}(x)$, $g_{\mu\nu}(\sigma)$ and $x^M(\sigma)$ around this
background gives the fluctuations $H_{MN}, h_{\mu\nu}, f^\mu$ and
$y^i$,  
\bea 
&&G_{MN}=\eta_{MN}+H_{MN}(x)\,,\qquad 
g_{\mu\nu}=\eta_{\mu\nu}+ h_{\mu\nu}(\sigma) \,,\\[.2cm] 
&&x^\mu_{||}(\sigma)=\sigma^\mu+ f^\mu(\sigma)\,,\qquad\qquad
x^i_\perp(\sigma)=y^i_0+y^i(\sigma)\,.
\label{fluc}
\eea
To first order, $h(x(\sigma))=h(x_{||})$, {\it etc.}, and the
fluctuations are related by the linearized pullback equation
$h_{\mu\nu}(x_{||})= H_{\mu\nu}(x_{||},y_0) +\p_\mu f_\nu +\p_\nu
f_\mu$. Infinitesimal bulk and brane diffeomorphisms transform the
fluctuations as,  
\be
\begin{array}{rlccc}
\delta x^M=\xi^M&:&  \quad
\delta_\xi H_{MN}=-2\p_{(M}\xi_{N)}\,,\,\, &
\delta_\xi h_{\mu\nu}=0\,, & \delta_\xi f^\mu=\xi^\mu\,,\\[.2cm]
\delta\sigma^\mu=\lambda^\mu&:& \quad
\delta_\lambda H_{MN}=0\,, &\delta_\lambda h_{\mu\nu} = -2
\p_{(\mu}\lambda_{\nu)}\,,\,\,& \delta_\lambda f^\mu=-\lambda^\mu\,.
\end{array}
\label{GT}
\ee  
The last equation arises since $x^M(\sigma)$ are scalars under brane 
diffeomorphism so that $\delta_\lambda f^\mu\equiv\delta_\lambda x^\mu
=-\lambda^\nu\delta^\mu_\nu$. The $f_\mu$ play a role similar to
St\"uckelberg fields in massive gravity.   

To solve the equations of motion, we use gauge invariant variables
instead of gauge fixing transformations (\ref{GT})\footnote{
  Customarily, one fixes static ({\it Monge}) gauge, $f^\mu=0$, on the  
  $x^\mu(\sigma)$ and harmonic gauge on $H_{MN}$.}. This disentangles
the spin $0$ and spin $2$ modes of the graviton and promotes $f_\mu$
to new gauge invariant variables. The bulk field $H_{MN}$ decomposes
as 
\be
H_{MN}=H^\perp_{MN}+\p_MA_N+\p_NA_M+\p_M\p_N\Phi+\frac{1}{d}
\,\eta_{MN} \, S
\label{H-giv}
\ee
where $\p^MH^\perp_{MN}=0, H^{\perp M}_M=0$, and $\p^MA_M=0$. Then,
$H^\perp_{MN}$ and $S$ are gauge invariant while $A_M$ and $\Phi$ give 
rise to the transformations of $H_{MN}$ in (\ref{GT}). The relevant
projection operators are listed in appendix A. The bulk
Einstein-Hilbert action becomes,
\bea
S^{bulk}_{EH}&=&-\frac{A}{4}\int d^dx \Big(\p^L H^{MN}\p_L H_{MN} -
\p^L H\p_L H +2 \p_L H\p_N H^{LN}-2\p_M H^{MN}\p^L H_{LN}\Big) 
\nn \\
&=& -\frac{A}{4}\int d^dx\left[\p^L H^{\perp MN}\p_L
  H^\perp_{MN}- \frac{(d-2)(d-1)}{d^2} \p^L S\p_L S\right]\,,
\label{S-bulk}
\eea
where gauge dependent terms drop out. To obtain the contribution to
the equation of motion from the second line, express the variations
$\delta H^\perp_{NM}$ and $\delta S$ in terms of $\delta H_{MN}$ using
the projection operators in Appendix A. The linearized
Einstein-Hilbert operator becomes, 
\be
({\cal E}_dH)_{MN}=\frac{1}{2}\Box_d H^\perp_{MN}
+\frac{d-2}{2d}\left(\p_M\p_N-\eta_{MN}\Box_d\right)S\,.
\label{EH-bulk}
\ee
Similarly, the induced brane field $h_{\mu\nu}$ is decomposed as
\be
h_{\mu\nu}=h^\perp_{\mu\nu}+\p_\mu a_\nu+\p_\nu a_\mu+
\p_\mu\p_\nu\phi+ \frac{1}{4}\,\eta_{\mu\nu}\, s \,,
\label{h-giv}
\ee
where $\p^\mu h^\perp_{\mu\nu}=0, h^{\perp \mu}_\mu=0$ and $\p^\mu
a_\mu=0$. $h^\perp_{\mu\nu}$ and $s$ are gauge invariant. The brane
Einstein-Hilbert action $S^{brane}_{EH}$ and operator $({\cal
  E}_4h)_{\mu\nu}$ can be read off from (\ref{S-bulk}) and
(\ref{EH-bulk}) for $d=4$, after appropriate field replacements. There
are also new gauge invariant variables based on $f^\mu$ and $y^i$ to
be introduced later in equations \ref{F} and \ref{Fi}.    

\subsection{Thick branes and ``blurred'' quantities}

Solving the equations of motion involves the massless scalar propagator
in $d$ dimensions,  
\be
G(x_{||}-x'_{||},x_\perp-x'_\perp) =-\int d^4 k\int d^n q \frac{e^{ik
    (x_{||}-x_{||}') +iq (x_\perp-x_\perp')}} {k^2+q^2-i\epsilon}\,.
\label{G}
\ee
$k_\mu$ and $q_i$ denote momenta parallel and transverse to the
brane. For two points restricted to lie on a zero thickness brane,
$x_\perp^i =x^{\prime i}_\perp=y_0^i$ and the $q$ integral in 
$G(x_{||}- x'_{||},0)$ diverges for $n>1$. Correspondingly, a bulk
field $S$ sourced by a brane localized source $\delta(x_\perp-y_0)
T(x_{||})$, {\it i.e.,} $S(x_{||}, x_\perp-y_0)=\int d^4 x'_{||}
G(x_{||}-x'_{||}, x_\perp-y_0) T(x_{||})$ also diverges as
$x_\perp\rightarrow y_0$. Hence, the restriction of such bulk fields
to the brane is ill defined. Essentially, a zero  thickness brane is a
point source in transverse directions leading to a
$|x_\perp-y_0|^{-(n-2)}$ divergence, as in the Coulomb or Yukawa
potentials. 

This problem is resolved by realizing that dynamics give branes a {\it
  form factor} and hence an effective thickness, as discussed in
\cite{Larus} for D-branes and in \cite{DGHS} for solitonic
branes. This is taken into account by replacing $\delta(x_\perp-y_0)$
by a brane {\it thickness profile function}, $P(x_\perp-y_0)$, with
normalization $\int d^nx_\perp P=1$. Then, the divergent brane
restriction $S(x_{||}, 0)$ of the bulk field is replaced by the
``blurred'' field $\la S\ra(x_{||})$, a weighted average over the
brane width,    
\be
\la S\ra(x_{||}) = \int d^nx_\perp P(x_\perp-y_0) S(x_{||},
x_\perp-y_0)\,, 
\label{blurrx}
\ee
With Fourier transform conventions $P(x_\perp-y_0)=\int d^nq\,\,\wt
P(q)e^{iq(x_\perp-y_0)}$, one gets the momentum space equation  
\be
\la \wt S\ra(k)=(2\pi)^n \int d^nq\,\,\wt P(q)\,\wt S(k,q)\,.
\label{blurr}
\ee

In the same way, the divergent brane restricted propagator $G(x_{||}-
x'_{||},0)$ gets replaced by its well defined thick brane analogue,  
$$
\la G\ra(x_{||}-x'_{||})=\int d^nx_\perp d^nx'_\perp\, P(x_\perp -y_0)
\,G(x_{||}-x'_{||},x_\perp-x'_\perp)\, P(x'_\perp-y_0)\,.
$$
Physically, this is the propagation amplitude from $x_{||}$ to
$x'_{||}$ located on the brane, with a small delocalization in the
transverse directions. This does not yet take into account physical
effects due to the presence of the brane. However, this construct will
contribute to the physical propagators and the interesting physics
associated with it will be discussed in the next section.

The corresponding momentum space expression (with $\wt P$ depending
only on $q^2$) is     
\be
\la \wt G\ra(k) =-(2\pi)^n\int d^nq \,\, 
\frac{[\wt P(q)]^2}{k^2 + q^2-i\epsilon} \,.
\label{tG}
\ee 
Again, for a zero-thickness brane, $\wt P=1$ and the expression
diverges for $n>1$. For thick branes, $\wt P(q)$
effectively implements a UV regularization at high $q$ through a
length scale associated with the brane width, $\omega$. Earlier, a 
non-zero brane width has been used to justify implementing a sharp 
cutoff $\sim 1/\omega$ on the $q$ integral \cite{Kiritsis,GS}, or 
solving the equations separately outside and inside the thick brane 
\cite{DGHS,Porrati}. Here, we will explicitly retain a smooth
profile function $\wt P$ through which the results depend on the  
brane width $\omega$. Such dependences are not affected by the actual
form of $P$, that encodes high energy effects and which will only
change the numerics. 

Since the normalized $P$ has mass dimension $n$, $\wt P(q)$ is
dimensionless. Then rotational invariance in the transverse space
means it only depends on the combination $q^2\omega^2$. This implies
(after simple manipulations in (\ref{tG})),
\be 
\la \wt G\ra(k)=\frac{1}{\omega^{n-2}}\, \Sigma^{\,-1}(\omega^2 k^2), 
\label{GSigma}
\ee 
depending on $k$ only through $u=\omega^2 k^2$. The expression
captures the form of divergence as $\omega\rightarrow 0$. Later,
$\Sigma$ will appear in the same way as a self-energy correction in
quantum field theory, hence the notation. Note that $w^n\la \wt G\ra$
has the correct dimension for a propagator.

The use of the brane effective width $\omega$ means that the theory is
valid for describing interactions of transverse gravitons with the brane 
as long as $q< 1/\omega$. Probing the brane at shorter scales is
meaningless in the low energy theory. To insure that {\it
  interactions} too cannot probe beyond the brane width, the effective
theory must also be restricted to $u=(k\omega)^2<1$, for momenta along
the brane.   

For the most part we do not need the functional form of $P$. But for
explicit calculations, a natural choice, motivated by
\cite{Antoniadis}, is the Gaussian form{\footnote{In contrast with a
    sharp cut-off on $q$, a Gaussian $\wt P(q)$ leads to a 
    non-negative $P(x_\perp-y_0)$, consistent with its interpretation
    as a thickness profile. The main results are not affected by these
    choices.}},  
\be
P(x_\perp)=\frac{1}{(\omega\sqrt{2\pi})^n}e^{-({x_\perp}/{2\omega})^2}\,,
\qquad
\wt P(q)=\frac{1}{(2\pi)^n}e^{-q^2\omega^2/2}\,.
\label{PG}
\ee
The profile function can also be discussed in a covariant setup, but
that is not needed here. 

\subsection{Bulk-brane relations} 

For a thick brane, the pullback equation relating bulk and brane
metric fluctuations becomes, 
\be
h_{\mu\nu}=\la H_{\mu\nu}\ra+\p_\mu f_\nu+\p_\nu f_\mu \,,
\label{lpb}
\ee
with only $x_{||}$ dependences. Note that fields intrinsically defined
on the brane are not affected by the ``blurring'' procedure
(\ref{blurrx}). Then, using (\ref{H-giv}) and (\ref{h-giv}) gives, 
\be
\Hp_{\mu\nu}=h^\perp_{\mu\nu}-\p_\mu F_\nu -\p_\nu F_\mu- 
\eta_{\mu\nu}\left(\frac{1}{d}\la S\ra - \frac{1}{4}s\right)\,.
\label{Hh}
\ee
The $F^\mu$ are new variables invariant under both bulk and brane
gauge transformations, 
\be
F_\mu=f_\mu+\la A_\mu\ra-a_\mu+\frac{1}{2}\p_\mu(\la\Phi\ra -\phi)\,.
\label{F}
\ee
The gauge dependent variables in $F_\mu$ do not appear in the action
(\ref{Sbig-I}). However, solving the equations of motion requires
either gauge fixing or using gauge invariant variables along with
(\ref{Hh}). Thus, in bulk-brane setups, gauge variant variables
survive through $F_\mu$ and contribute to the solutions for the brane
fields $s$ and $h^\perp$. The implication of this as a broken phase
realization of $4$-dimensional general covariance is discussed in
the last section.

\section{The Basic Brane Induced Gravity Model}

In this section, we review and further investigate the solutions
of the {\it basic} BIG model \cite{DGP,DG} based on bulk and brane
Einstein-Hilbert actions of (\ref{Sbig-I}) but without
$S^{brane}_\Lambda$ and $S^{brane}_\Omega$. First, we solve the
linearized equations of motion for the brane fields and review the
tachyon/ghost problem \cite{DR}. The propagator is then analyzed in
depth, extracting a mass, a decay width and a modulation of the
gravitational coupling. The suppression of graviton decay into the
bulk is explained with the help of the optical theorem. The reader not
interested in the details of solving the equations of motion can
directly jump to the solutions (\ref{s}) and (\ref{hp}).

\subsection{Equation of motion and solutions}

For the basic BIG model, the linearized equation of motion for
$H_{MN}$ expressed in terms of Einstein-Hilbert operators
(\ref{EH-bulk}) and for thick branes, takes the form 
\bea
A({\cal E}_dH)^{MN}+\,P(x_\perp-y_0)\left(B({\cal E}_4h)^{\mu\nu}\,\,
+\frac{1}{2}T^{\mu\nu}\right)\delta^M_\mu\delta^N_\nu=0\,. 
\label{eomHAB}
\eea
In our momentum space conventions, it becomes\footnote{The $(\mu\nu)$
  components alone can determine the brane fields, but it is
  more convenient to use all equations.}(with $p^M=\{k^\mu, q^i\}$),  
\bea
&&\!\!\!\!\!\!\!\!\!\!\!\!\!\!\!\!\!\!\!\!\!\!
-A\left[(k^2+q^2)\wt H^{\perp MN}+\frac{d-2}{d}\left(p^M
  p^N- \eta^{MN}(k^2+q^2)\right)\wt S \right]_{(k,q)} \nn\\ 
&&\qquad -\wt P(q)\left[B\left(k^2 \wt h^{\perp \mu\nu}+
\frac{1}{2}\left(k^\mu k^\nu-\eta^{\mu\nu}k^2\right)\wt s \right)
-\wt T^{\mu\nu}\right]_{(k)}
\delta^M_\mu\delta^N_\nu =0\,. 
\label{eom}
\eea
This is to be combined with the surface equation (\ref{Hh}).
In the following, the $i\epsilon$ terms in the bulk propagator 
are not always written explicitly, but are finally included  
in the blurred propagator (\ref{tG}).

{\it The $\eta_{MN}$-trace} of (\ref{eom}) determines $\wt S$ in
terms of $\wt s$,
\be
\wt S(k,q)=-\frac{d}{A(d-1)(d-2)}\,\,\frac{\wt P(q)}{k^2+q^2-i
\epsilon}\,\,\left(\wt T^\mu_\mu+\frac{3}{2}B\,k^2\,
\wt s\right)_{(k)} \,.
\label{tS}
\ee
Restricting to the brane by using (\ref{blurr}) and the expression for 
$\la\wt G\ra$ in (\ref{tG}), one gets,  
\be
\la\wt S\ra(k)=\frac{d}{A(d-1)(d-2)}\,\,\la \wt G \ra
\,\,\left(\wt T^\mu_\mu+\frac{3}{2}B\, k^2\, \wt s \right)_{(k)}\,.
\label{blS}
\ee 
For later convenience, we express $\wt S(k,q)$ in terms of $\la\wt
S\ra(k)$, 
\be
\wt S(k,q)=-\frac{\wt P(q)}{k^2+q^2}\,\, 
\frac{1}{\la\wt G\ra}\,\,\la\wt  S\ra \,.
\label{SbrakS}
\ee

{\it The $(\mu, i)$ components} give (since ``blurring'' now involves 
integrating over an odd function of $q^i$),
\be
\wt H^{\perp\mu i}(k,q)=-\frac{d-2}{d}\,\,\frac{k^\mu q^i}{k^2+q^2}
\,\,\wt S(k,q)\,,\quad\Rightarrow\quad
\tHp^{\mu i} (k)=0 \,.
\label{Hmui}
\ee
However, note that using (\ref{SbrakS}), and the transversality of
$H^\perp$, one gets,
\be
\la q_i\wt H^{\perp\mu i}\ra(k)=-k_\nu\tHp^{\mu\nu}=
\frac{d-2}{d}\,\frac{\la\wt g\ra}
{\la\wt G\ra}\, k^\mu \la\wt S\ra \,.
\label{qvecH}
\ee
where we introduce a new function $\la\wt g\ra(k)$ (that
will drop out of most expressions),
\be
\la\wt g\ra=(2\pi)^n\int d^nq \frac{q^2\wt P^2}{(k^2+q^2)^2} \,.
\label{gbrak}
\ee

{\it The $(i,j)$ components} give, 
\be
\wt H^{\perp ij}(k,q)=-\frac{d-2}{d}\,\,
(\frac{q^iq^j}{k^2+q^2}-\eta^{ij}) \,\,\wt S(k,q)\,.
\label{HijS}
\ee
From this one can immediately see that $\la q_i\wt H^{\perp ij}\ra=0$
and, for $i\neq j$, $\la \wt H^{\perp ij}\ra=0$. However,
\be
\la\wt H^{\perp j}_j\ra(k)=\frac{d-2}{d}\,\,
\left(\frac{\la\wt g\ra}{\la\wt G\ra}+n \right)\,\,
\la\wt S\ra \,,
\qquad
\la q_iq_j\wt H^{\perp ij}\ra(k)=-\frac{d-2}{d}\,\, k^2\,\,
\frac{\la\wt g\ra}{\la\wt G\ra}\,\,\la\wt S\ra \,.
\label{trHqqH}
\ee

{\it The surface equation} (\ref{Hh}), on taking a trace and a
double divergence, gives
\be
\la \wt H^{\perp\mu}_\mu\ra =-2i k^\mu\wt F_\mu 
-\frac{4}{d}\la\wt S \ra + \wt s \,, \qquad
k^\mu k^\nu\tHp_{\mu\nu}=
-k^2\left(2i k^\mu\wt F_\mu+\frac{1}{d}\la \wt S\ra-\frac{1}{4}\wt s
\right) \,.
\label{TrDD}
\ee
Since $\la \wt H^{\perp\mu}_\mu\ra =-\la \wt H^{\perp i}_i\ra$ and  
$k^\mu k^\nu\tHp_{\mu\nu}=\la q_iq_j\wt H^{\perp ij}\ra(k)$, on using
(\ref{trHqqH}), (\ref{TrDD}) become,
\be
2ik^\mu F_\mu=\left[(d-2)(\frac{\la\wt g\ra}
{\la\wt G\ra}+n )-4\right]\frac{\la\wt S\ra}{d}+\wt s \,,
\quad
2ik^\mu F_\mu=\left[(d-2)\frac{\la\wt g\ra}{\la\wt G\ra}
-1\right]\frac{\la\wt S\ra}{d}+\frac{\wt s}{4} \,.
\ee  
Eliminating  $k_\mu\wt F^\mu$ gives a relation between $\la\wt S\ra$
and $\wt s$ as, 
\be
\frac{(d-1)(d-5)}{d}\,\la\wt S\ra =-\frac{3}{4}\,\wt s \,.
\label{Ss}
\ee
Substituting back in the expression for $k_\mu\wt F^\mu$ determines
it in terms of $\wt s$. Combining this with the divergence of
(\ref{Hh}) and then using (\ref{qvecH}) leads to the solution for the
$\wt F^\mu$ fields,  
\be
\wt F^\mu=-\,\frac{i}{4}\frac{k^\mu}{k^2}\,U\,\wt s\,,\quad{\rm
  where}\,,\quad U=\frac{1}{2}\frac{d-2}{(d-5)(d-1)}\left(n-3\,
\frac{\la\wt g\ra}{\la\wt G\ra}\right) \,. 
\label{Fmu} 
\ee
Finally, combining (\ref{Ss}) with (\ref{blS}) gives the solution
(\ref{s}) below for $\wt s$ in terms of $T=T^\mu_\mu$.

{\it The $(\mu,\nu)$ Components} of (\ref{eom}) can be solved for   
$H^\perp_{\mu\nu}$. One can then compute $\la\wt H^\perp_{\mu\nu}\ra$
using (\ref{SbrakS}). Using (\ref{Hh}), this becomes an equation
for $\wt h^\perp_{\mu\nu}$ in terms of $\wt S$, $\wt s$ and $F^\mu$
all of which are known in terms of $\wt T_{\mu\nu}$. Thus one has the
final solutions, 
\bea
&\wt s(k)=-\ds\frac{2}{3B}\,\frac{1}{k^2+\frac{A}{B}\,
\frac{d-2}{2(d-5)}\,\la\wt G\ra^{-1}}\,\wt T\,,  & 
\label{s} 
\\[.3cm]
&\wt h^\perp_{\mu\nu}=\ds\frac{1}{B}\,\frac{1}{k^2-\frac{A}{B}\,
\la\wt G\ra^{-1}}\left(\wt T_{\mu\nu}-\frac{1}{3}(\eta_{\mu\nu}-
\frac{k^\mu k^\nu}{k^2})\,\wt T\right) \,.& 
\label{hp}
\eea 

From this one can directly read off the gauge independent parts of the
Greens functions $G_{\mu\nu\mu'\nu'}^{(0,2)}$, or the $4$-dimensional
gauge invariant amplitude,   
\bea
&& \int d^4x\int d^4x' T^{\mu\nu}(x)G_{\mu\nu\mu'\nu'}(x-x') 
          T^{\mu'\nu'}(x') \nn \\ 
&& \qquad\qquad\qquad\qquad = \int d^4x T^{\mu\nu}(x) h_{\mu\nu}(x)
=\int d^4x \left(T^{\mu\nu} h^\perp_{\mu\nu} +\frac{1}{4}T\, s\right)\,.
\label{gia}
\eea
If needed, the $h_{\mu\nu}$ in a given gauge can be constructed by
adding gauge transformations,
$h_{\mu\nu}=h^\perp_{\mu\nu}+\frac{1}{4}\,\eta_{\mu\nu}\,s+\p_\mu\xi_\nu+
\p_\nu \xi_\mu$, and solving the gauge conditions for the $\xi_\mu$.

One can immediately draw a number of conclusions, mostly known in the
literature, based on the structure of $s$ and $h^\perp$ solutions: 

(1) {\bf Zero thickness limit:} Note that in the absence of extra
dimensions, $\la\wt G\ra_{n=0}=-1/k^2$, leading to the solutions for
ordinary Einstein-Hilbert gravity, 
\be
(\wt h^\perp_{\mu\nu})_{(n=0)}=\frac{1}{A'}\,\, \frac{1}{k^2}
\left(\wt T_{\mu\nu}-\frac{1}{3}(\eta_{\mu\nu} -
\frac{k^\mu k^\nu}{k^2})\wt T\right)\,,\qquad
(\wt s)_{n=0}=-\frac{2}{3A'}\,\frac{1}{k^2}\,\wt T\,.
\label{hsd=4}
\ee 
  Also $\la\wt g\ra_{n=0}=0$ gives $(\wt F^\mu)_{n=0}=0$. The BIG
  results (\ref{s}),(\ref{hp}) differ from this only in the $\la\wt
  G\ra^{-1}$ dependent terms. Now, while the brane width $\omega$ was
  introduced to keep $\la\wt G\ra$ finite for $n>1$, the final
  solution is well defined for $\omega\rightarrow 0$. Then, $\la\wt
  G\ra$ diverges (from (\ref{GSigma})), resulting again in
  (\ref{hsd=4}). Hence, for $n>2$, modifications of $4$-dimensional
  gravity arise only for thick branes. Since the spin $2$ and spin $0$
  parts of the propagator in BIG have the same coefficients as in
  ordinary gravity (except $d=5$ where $\wt s=0$), there is no vDVZ
  discontinuity \cite{vDVZ1,vDVZ2,vDVZ3} in the $\omega\rightarrow 0$
  limit.  

 (2) {\bf Tachyon problem:} For $n>2$, $\la\wt G\ra(k^2=0)$ is finite
  and contributes a hard mass to the propagators. The scale of the
  mass is set by $A\omega^{n-2}/B$ and the gravitational coupling is
  given by $G_N\sim 1/B$. Thus, for phenomenological reasons, $B$
  should be large and $A\omega^{n-2}$ should be small. But $\la\wt
  G\ra^{-1}$ appears with opposite signs in the denominators,
  indicating that if $h^\perp$ has a healthy mass, then $s$ will be
  tachyonic or vice versa \cite{DR}. For the conventional choice of
  $B>0$ and $A>0$, $h^\perp$ cannot have a tachyonic pole (since for
  any $k^2>0$, (\ref{tG}) gives $\la\wt G\ra(k)<0$). Then, $s$ is
  tachyonic.
  
  3) {\bf Ghost problem:} It is well known \cite{vDVZ1,VN,Nunes} that
  a propagator for $h_{\mu\nu}$ with the tensor structure of ordinary
  massless gravity in $4$-dimensions, is ghost free only for zero
  graviton masses. Thus the non-zero masses and the $-2/3$ factor in
  the $\wt s$ solution makes the tachyonic spin $0$ mode also a ghost
  \cite{DR}. The presence of this tachyonic ghost in brane induced
  gravity has so far hampered its further development. One of our
  purposes in this paper is to shed light on the origin of this ghost
  mode. 

  4) {\bf Unstable gravitons:} Ignoring the tachyonic ghost $\wt s$
  field, the massive spin $2$ graviton $h^\perp$ on the brane is not a
  stable particle but a resonance state. Technically, this is because
  $\la\wt G\ra$ has a branch cut from $k^2=0$ to $-\infty$ due to the
  continuum of Kaluza-Klein modes resulting from an uncompactified
  bulk (\cite{DGS1,DGS2}). For very small masses, the life time is
  very large. In the next subsection we will explain the origin of the 
  decay and its {\it suppression} using standard quantum field theory
  and then carry out a detailed analysis of the propagator.

  5) {\bf String theory limit:} The string theory computation of the
  BIG action in \cite{Corley,Ardalan} obtained $B<0$, making $h^\perp$
  the tachyonic ghost. But in this 
  setup, in terms of the string coupling $g_s$, $A\sim 1/g_s^2$ and
  $B\sim 1/g_s$. Assuming $\omega\sim g_s^0$, weak string
  coupling corresponds to large masses, beyond the validity
  limits of the theory. Even for D3-branes where $n=6$ and
  $\omega^4\sim g_s$ is a more natural choice, masses are
  $g_s$-independent and at string scale. In either case, for $k^2$ 
  values within the validity range of the theory, the propagator is
  given by $-\la\wt G\ra(k)/A$, avoiding the tachyon/ghost problems,
  and gravity is essentially higher dimensional. However, the orbifold
  based construction of \cite{Antoniadis} that attempts to derive BIG
  from string theory with the phenomenologically interesting parameter
  ranges, does not evade the tachyon/ghost problem in this manner.

In the rest of this section we concentrate on the healthy spin $2$
mode, returning to the origins of the ghost mode in the next section. 
  
\subsection{Graviton decay from optical theorem}

A main feature of BIG, analyzed around flat background, is that
$h^\perp$ is a massive {\it unstable} graviton in $4$ dimensions due
to the possibility of gravitons escaping into the bulk
\cite{DGP,DGS1,DGS2}.  However, the graviton lifetime must be large
for phenomenological reasons. Below, we show that the suppression of
graviton decay can be explained using general field theory
arguments. The detailed structure of the propagator is analyzed in the
next subsection.

The bulk propagator $G(x-x')$ (\ref{G}) gives the amplitude for
graviton propagation from any $x$ to any $x'$ (ignoring the tensor
structure and for canonically normalized fields, so no $1/A$ factor).
But $\la G\ra(x_{||}-x'_{||})$ is constructed to describe a restricted
propagation between two points with their $x_\perp$ coordinates
delocalized over the width $\omega$ of a thin region that would be
occupied by the brane. Of course, there is a finite probability
$\sigma_{escape}$ that gravitons emitted at $x_{||}$ within the thin
region, end up somewhere in the bulk. Such processes are not 
described by the restricted propagator $\la G\ra$ and appear to it as
decay channels, thus giving it an imaginary part. The relationship is
quantified by the optical theorem where, in terms of the corresponding 
amputated $2$-point function $\omega^{-n}\la\wt G\ra^{-1}$ (with
$\omega$ factors restoring the correct dimensions for the propagator),   
$$
2\, Im\,(\omega^{-n}\la\wt G\ra^{-1})\sim \sigma_{escape} \neq 0 \,. 
$$
This is a geometric result that holds for any region of space
that would be occupied by a thick brane and applies to gravitons not
yet dynamically affected by a physical brane.   

Now, with a physical brane in place and after taking gravitational
dynamics on the brane into account, we obtained a brane-to-brane
propagator $\wt G_{bb}=-(k^2-\frac{A}{B}\la\wt G\ra^{-1})^{-1}$ in
(\ref{hp}), but now written for the canonically normalized brane field   
and with suppressed tensor structure. The stability of the
corresponding state is again encoded in the imaginary part of the
amputated $2$-point function which now gives,  
\be 
2 Im\,\,\wt G_{bb}^{-1} = 2 \frac{A}{B}\,Im\,\la\wt G\ra^{-1} 
\sim\frac{A\omega^n}{B}\, \sigma_{escape}\,.
\ee
Hence, graviton decay on the brane is still due to escape into the
bulk, but with an amplitude suppressed by a factor $A\omega^n/B$ as
compared to the unhindered escape in the absence of brane
gravitational dynamics. This is a way of seeing that a large
brane Einstein-Hilbert term confines gravitons to the brane by
suppressing their escape into the bulk.

Finally, for completeness, one can also compute the brane-to-bulk
propagator,  
$$
\wt G_{Bb}(k,q)=\frac{-\wt P(q)\,\la\wt G\ra^{-1}}{k^2+q^2}\,
\left[\frac{1}{B}\frac{1}{k^2-\frac{A}{B}\la\wt G\ra^{-1}}\right]\,. 
$$
The brane-to-bulk decay probability $\sigma_{brane\rightarrow bulk}$
can be constructed out of this after proper amputations (using
$G_{bb}$ for the brane leg and $G$ for the bulk leg). One can then
verify the optical theorem for the physical propagator, 
$$
2Im\,\wt G_{bb}^{-1}\sim\sigma_{brane\rightarrow bulk}
$$
In this sense, the verification of the optical theorem on the brane
requires using the brane-to-bulk propagator, since the brane theory
alone is not unitary.  

\subsection{Graviton mass, decay width and coupling}

Let us investigate the properties of the spin $2$ graviton $h^\perp$
(\ref{hp}) in more detail. The physics of the corresponding
brane-to-brane propagator follows from the structure of $\la\wt
G\ra(k^2-i\epsilon)$. We will see that the propagator has the standard
form for a massive unstable particle in QFT. More importantly, for
small masses it is amenable to the standard particle physics
approximation methods.  

To evaluate $\la\wt G\ra$ (\ref{tG}) we use the Gaussian form
(\ref{PG}) for $\wt P(q)$. Doing the angular integrals in $q$-space
gives, 
\be
\la\wt G\ra = -2N\int dq\,q^{n-1}\frac{e^{-q^2\omega^2}}{k^2+q^2
-i\epsilon}\,,\qquad  N^{-1}=(4\pi)^{\frac{n}{2}}\,\Gamma(\frac{n}{2})\,.
\ee   
First, let us consider the {\it even} $n$ case and set
$2m=n-2$ (the result for {\it odd n} is given later). After some
manipulations,  
\be
\la\wt G\ra(k)=-\,\frac{(-1)^m\,N}{\omega^{n-2}}\,\, 
u^m \,\left(\frac{\p}{\p u}\right)^m\Big(e^u\,E_1(u-i\epsilon)\Big)\,,
\ee   
where $u=\omega^2k^2$ and we have used the notation \cite{Abramowitz},
\be
E_1(u-i\epsilon)=\int_{u-i\epsilon}^\infty dt \frac{e^{-t}}{t} \,.
\ee
Since $\p_u E_1(u)=-e^u/u$, for $\Sigma^{-1}=\omega^{n-2}\la\wt G\ra$
one gets   
\be
\Sigma^{-1}(u-i\epsilon)  =-(-1)^m\,N\, 
\left[u^m\,e^u\,E_1(u-i\epsilon)+\sum_{r=1}^{m}(-1)^r\,(r-1)!\,u^{m-r} 
\right] \,.
\label{Seven-n}
\ee   
The corresponding expression for {\it odd n} is given by equation
(\ref{Sodd-n}) below.

Remember that the $i\epsilon$ prescription tells us how to handle the
$k^2+q^2=0$ poles in (\ref{tG}). Now, after performing the
$q$-integrations, the same prescription will dictate the correct
$\epsilon\rightarrow 0$ limit. To see what happens, note that the
function $E_1(z)$ is the analytic continuation, to the complex plane,
of the {\it exponential integral} $Ei(-u)$ defined, for real $u$, by 
\be
Ei(-u)=-\int_{u}^\infty dt \frac{e^{-t}}{t} \,,
\ee
$Ei(-u)$ is defined over the entire real line, but $E_1(z)$ has a
branch cut from $z=0$ to $z=-\infty$ along the negative real axis. The
crucial point is that as $\epsilon\rightarrow 0$, we approach the
branch cut from {\it below} and
\be
\lim_{\epsilon\rightarrow 0} E_1(u-i\epsilon)=- E_i(-u)+i\pi\,\theta(-u)\,,
\ee
where, $\theta(-u)$ is the unit step function. The imaginary part is
the origin of the complex pole in the propagator, and hence of
graviton decay \footnote{Approaching the branch cut from {\it above}
  (corresponding to $E_1(u+i\epsilon)$), would give $-i\pi$ which does
  not lead to decay. So the correct sign is dictated by the
  $i\epsilon$ prescription in the bulk propagator.}. Writing
$\Sigma^{-1}=a+ib$, the real and imaginary parts are given by
\bea
&& a(u)=(-1)^m\,N\,\left[u^m\,e^u\,Ei(-u)
-\sum_{r=1}^{m}(-1)^r\,(r-1)!\,u^{m-r} \right] \,, 
\label{a}\\
&& b(u)= -\pi\, N\, |u|^m\,e^{-|u|}\,\theta(-u) \,.
\label{b}
\eea
Finally, in terms of $\Sigma=\Sigma_1+i\Sigma_2$, the brane-to-brane
propagator in the $h^\perp$ solution (\ref{hp}) is (with canonical
field normalization and suppressed tensor structure),  
\be
\wt G_{bb}=\frac{-1}{k^2-\frac{A\,\omega^{n-2}}{B}\,\,
\Big[\,\Sigma_1(\omega^2k^2)+ i \Sigma_2(\omega^2k^2)\,\Big]} \,,
\label{Gbb}
\ee
The above analysis then gives, 
\be
\Sigma_1(u)=\frac{a}{a^2+b^2}\,,\qquad\qquad 
\Sigma_2(u)=\frac{-b}{a^2+b^2}\,\geq 0 \,.
\label{sigma12}
\ee

This propagator has the familiar quantum field theory form where the
analogue of $\Sigma$ arises due to self-energy corrections. It has a
complex mass pole and, strictly speaking, the graviton {\it mass} and
{\it decay width} are given by the real and imaginary parts of the
complex pole. In practice, determining the location of the complex
pole is not easy. But since the determination of mass anyway becomes
somewhat arbitrary due to the finite life-time, an approximation can
be made provided, near the mass pole, $|\Sigma_1|$ is much larger than
$|\Sigma_2|$. In such cases, a mass can be defined as the pole
position with only $\Sigma_1$ present, while $\Sigma_2$ gives the
decay width (see, for example, \cite{Peskin, Brown}). This is often a
valid approximation in perturbative quantum field theory. Fortunately,
one can show that it also holds here for small enough masses: For
$m\geq 1$, as $u\rightarrow 0$, $\Sigma_2$ vanishes as $u^m$ while
$\Sigma_1$ goes to a constant $-1/[N(m-1)!]$. Hence, for a small
enough mass, at the mass pole $M^2=-u/\omega^2$, one has
$|\Sigma_1|>>|\Sigma_2|$, as wished.

Then, in this approximation the mass $M^2=-k^2$ is determined, on
ignoring $\Sigma_2$, by    
\be
\left[\frac{B}{A\omega^{n}}\right]\,\omega^2\,k^2=
\Sigma_1(\omega^2k^2)\,.  
\label{mass}
\ee
For $n>2$, $\Sigma_1(0)=-[N(m-1)!]^{-1}<0$. So, by continuity,
$\Sigma_1(\omega^2k^2)<0$ for small arguments in the neighborhood of
zero. Then the graviton is non-tachyonic, $M^2>0$. More concretely,
this is visible in Fig.1(a) where the right hand side of the above
equation is plotted against $u=\omega^2k^2$ for $n=6,5,\cdots, 3$
({\it odd n} curves are based on (\ref{Sodd-n})) and the left hand
side is plotted for two different slopes $B/A\omega^n$, with values
chosen only for illustrative purposes. The intersections determine the
masses $M^2=-u/\omega^2$, with larger slopes resulting in smaller
masses. The plots also show that $\Sigma_1$ is monotonic for $n>2$ and
the propagator has a single mass pole. In Fig.1(b) we plot $\Sigma_2$
{\it vs} $\omega^2k^2$. Comparing curves for the same $n$ in both
plots again shows that for very small masses,
$|\Sigma_2(-\omega^2M^2)|<<|\Sigma_1 (-\omega^2M^2)|$, justifying the
approximation (the curves for $n=1,2$ look rather different and are
not shown here).
\begin{figure}[!ht]
\begin{tabular}[c]{cc}
\includegraphics[width=2.8in]{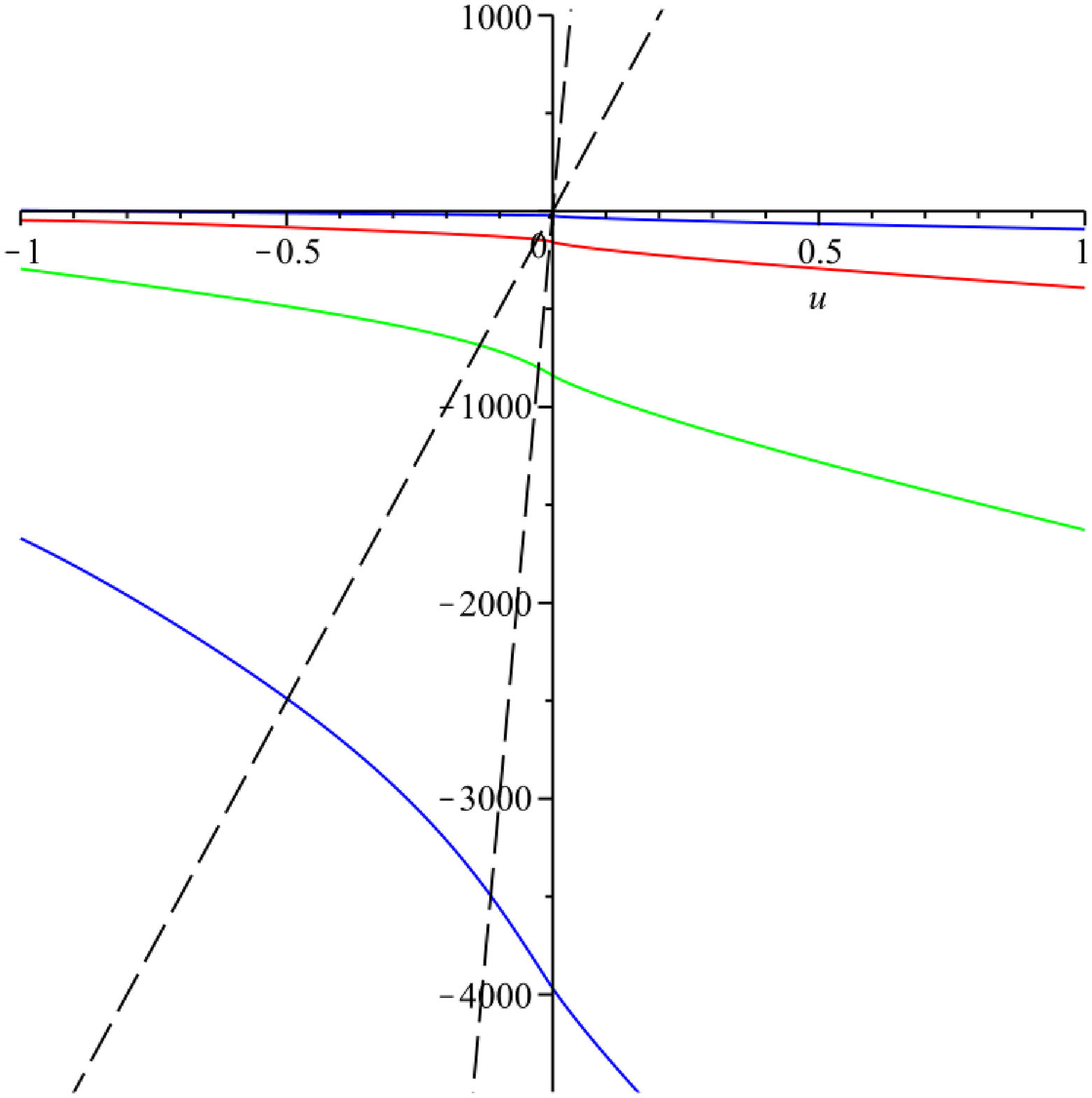}
&
\includegraphics[width=2.8in]{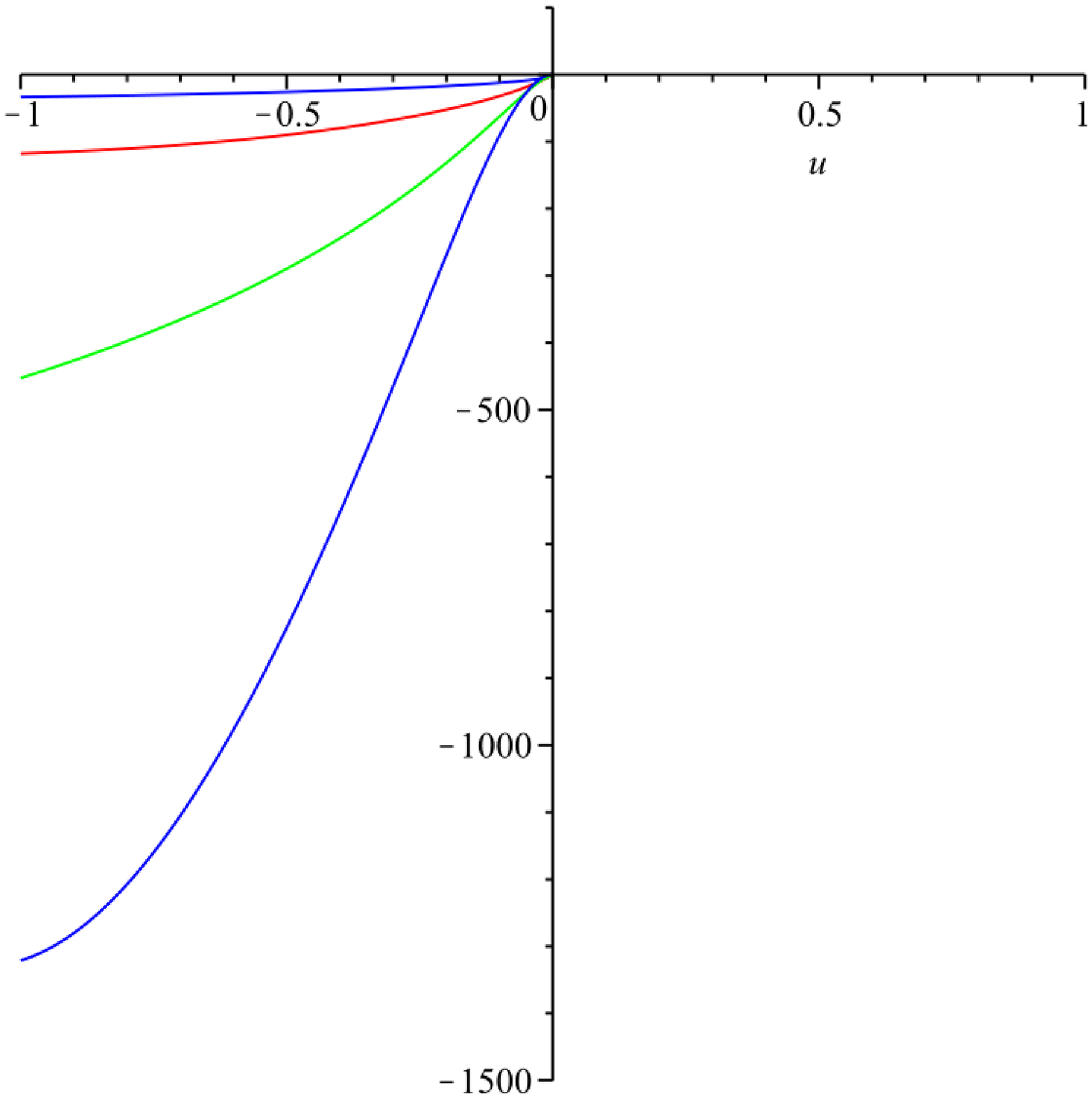}\\
(a) Determination of Mass & (b) Decay Width
\end{tabular}
\caption{\it{Behaviour of $\Sigma_1(u)$ (left) and $\Sigma_2(u)$
    (right) for codimensions $n=6$ (the lowest curves) through $n=3$
    (close to the x-axis)}}
\end{figure}

Now, in terms of the mass as determined above, the propagator has 
the standard form,  
\be
\wt G_{bb}=-\,\frac{Z(k)}{k^2+M^2-i\,\frac{A\omega^{n-2}}{B}\,Z(k)\,
\Sigma_2(k)}\,,
\label{GbbZ}
\ee
where, 
\be
Z^{-1}(k)= 
1-\frac{(A\omega^{n-2}/B)\,\Sigma_1(\omega^2k^2)+M^2}{k^2+M^2}\,.
\label{Z}
\ee
This will look more familiar when $\Sigma_1(u)$ is Taylor expanded 
around $u=-\omega^2M^2$.

Note that $Z$ retains a momentum dependence which amounts to a
momentum dependence of the effective gravitational coupling,
\be
G_N^{eff}(k)=Z(k)/B\,.
\label{GNk}
\ee
This momentum dependence can be expressed entirely in terms of the
dimensionless parameters $u$ and $A\omega^n/B$. In particular, for
time independent sources ($k^2>0$), $\Sigma_2=0$ and this is the only 
modification besides the graviton mass.

A simple analysis of $Z(k)$ shows that at large momenta, $|k^2|>>M^2$,
$Z\sim 1$. But for small momenta, $|k^2|\sim M^2$, on Taylor expanding
around the mass pole given by (\ref{mass}), one gets $Z<1$. Hence the
gravitational coupling is weaker at large distances, but the variation
is not drastic. This is over and above the effects due to graviton
mass and decay width.

Now, comparing (\ref{GbbZ}) to the relativistic Breit-Wigner form near
the mass pole, the decay width is identified as       
\be
\Gamma=\frac{A\omega^{n-2}}{M\,B}\,Z(u)\,\Sigma_2(u)\Big\vert_{u=
-\omega^2M^2} = -MZ(u)\frac{\Sigma_2(u)}{\Sigma_1(u)}
\Bigg\vert_{u=-\omega^2M^2}> 0 \,.
\label{decay}
\ee
Once again, note that it is the $i\epsilon$ prescription in the bulk
propagator that leads to the correct sign for the decay width. 

For odd $n$ the analysis proceeds along similar lines, except that the 
exponential integral is replaced by the {\it error function} and, with
$2m'=n-1$, we get,
\bea
\Sigma^{-1}(u-i\epsilon)&=&-(-1)^{m'}\,N\pi\, 
\Big[u^{m'-\frac{1}{2}}\,e^u\,\left(1-erf(\sqrt{u-i\epsilon})\right)
\nn\\
&&\hspace{4cm}  +\,\frac{1}{\sqrt{\pi}}\sum_{r=1}^{m'}(-1)^r\,
\frac{(2r-3)!!}{2^{r-1}}\,u^{m'-r}\Big]\,.
\label{Sodd-n}
\eea
\begin{wrapfigure}{r}{2.9in} 
\includegraphics[width=2.9in]{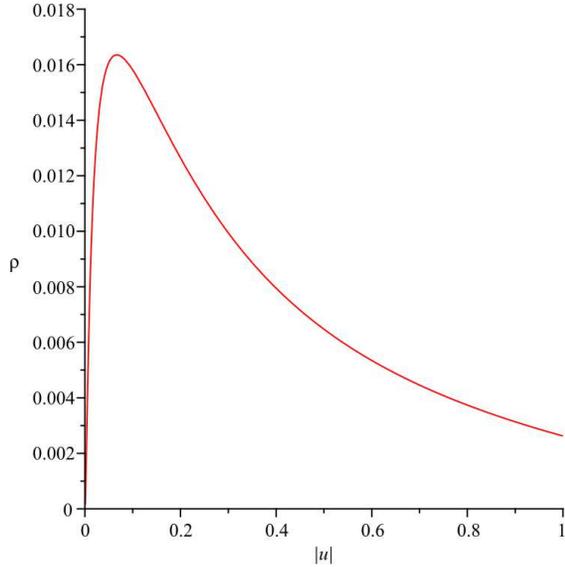}
\caption{\it{A Sample Spectral Density Function}} 
\end{wrapfigure}
In fact, this expression has been used to plot the curves for $n=3,5$
in Fig.1(a) and Fig.1(b). A unified description of both even and odd
$n$ is given in terms of the {\it Whittaker functions},  
$$
\Sigma^{-1}=-\,\frac{u^{\frac{n}{4}-1}e^{\frac{1}{2}u}}
{(4\pi)^{\frac{n}{2}}} \,\,W_{-\frac{n}{4},\frac{n-2}{4}}(u-i\epsilon)\,.
$$  

For completeness, in Fig.2, we show a representative plot for the
spectral density function,
$$
\rho(s)=-\frac{1}{\pi}Im G_{bb}(k^2=-s)\,,
$$
to show that it too has the standard form for an unstable particle, 
again with the $A\omega^n/B$ value chosen only for the purpose of 
illustration.

To summarize, we have shown that although the origin of the
``self-energy'' contribution, $\Sigma$, to the propagator in the BIG
setup is very different from its origin in perturbative quantum field
theory, the resulting unstable massive particles behave in very
similar ways and are amenable to the same approximation. Hence, the
modification of the spin $2$ graviton propagator (\ref{hp}), contained
in $\la\wt G\ra^{-1}$, breaks up into $3$ contributions: a hard mass
$M$ given by (\ref{mass}), a decay width $\Gamma$ (\ref{decay}) and a
momentum dependent Newton ``constant'' (\ref{GNk}).

\section{Screening of $\Lambda$ and the Origin of Tachyon/Ghost
  Problem}  

The possible resolution of the cosmological constant problem is the
main reason for interest in $n>2$ brane induced gravity, despite its
ghost/tachyon issues. Support for this expectation comes, as discussed
in \cite{DGS1,DGS2}, from the classical solutions of $d$-dimensional
gravity \cite{Gregory:1995qh,Charmousis} that describe a 3-brane with
an ADM mass density corresponding to a non-zero brane tension (which
is the same as the brane cosmological constant). To be precise, {\it
  the classical solutions describe this setup at distances far away
  from the location of the brane core}. The interpretation  is
that, for codimensions $n>2$, a brane cosmological constant $\Lambda$
can curve directions transverse to the brane rather than those
parallel to it, keeping the brane metric
flat\footnote{\cite{DGS1,DGS2} also consider time dependent solutions
  but these do not show up in our linearized analysis and are not
  discussed here.}. Thus the effect of $\Lambda$ is absorbed by bulk
curvature and it does not curve the brane worldvolume to a de Sitter
spacetime.

Now, the presence of a brane Einstein-Hilbert term in the
action does not affect this classical solution. But for a large enough
coefficient, it is expected to dictate the dynamics of metric
perturbations $h_{\mu\nu}$ around the classical solution. This is why
one expects $n>2$ BIG models to describe effectively $4$-dimensional
gravity with suppressed $\Lambda$ (The $n=2$ model is somewhat special
and has been analyzed in some detail in
\cite{Corradini,Kaloper1,Kaloper2}).

Unfortunately, implementing the above scenario is not as
straightforward.  In reality, to study metric perturbations
$h_{\mu\nu}$, one needs classical solutions that are valid in the
vicinity of the brane (at least with some effective thickness
included). The existing solution \cite{Gregory:1995qh,Charmousis} is
not adequate for this purpose since the energy momentum tensor that
sources it near the brane core, does not correspond to a brane
$\Lambda$ source.

Hence, in practice, in BIG with codimension $n>2$, the analysis of
metric perturbations has mostly been performed in a flat background,
instead of one sourced by $\Lambda$ (even though the latter is used to
argue for the suppression of $\Lambda$). It is also in the flat
background analysis that one encounters the tachyonic ghost problem
reviewed in the previous section. One may conjecture that this problem
is an artifact of the flat approximation and would go away once a full
fledged analysis is performed around a curved background sourced by
$\Lambda$. However, in the absence of a classical solution adequate
for such analysis, here we point out that the flat background analysis
of the previous section is, by itself, sufficient to study the
response of gravity to a small brane cosmological constant and exhibit
its suppression. This also has ramifications for what to expect from a
curved background analysis for the resolution of the ghost issue (as
will be detailed below). The healthy spin $2$ part of the graviton, on
which we have concentrated so far, is not relevant here.  Below we
return to the unhealthy scalar mode to find that,
\begin{itemize}
 \item The ``tachyonic'' mass is actually healthy and is instrumental
   in filtering out $\Lambda$ in the flat background approximation.
   It arises, at least partly, as an artifact of the flat
   approximation and as a substitute for the background curvature.
\item For phenomenologically interesting values of parameters
  (corresponding to a small graviton mass), the ghost nature is more
  intractable and {\it will most likely persist} even when metric
  perturbations are analyzed in the background curved by $\Lambda$,
  unless one ends up with massless gravity on the brane in a fully
  non-linear background.
\item The $4$-dimensional Newton constant and other couplings are
  modified by the screened $\Lambda$, consistent with
  \cite{G-rev,DGS1,DGS2}. This may help in avoiding a very low bulk
  gravity scale $A$, but may introduce hierarchy problems for other
  couplings. 
\item Finally we emphasize that BIG and massive Fierz-Pauli gravity
  (as studied in \cite{DHK}) use different mechanisms to respond to 
  $\Lambda$, in spite of a superficial similarity.
\end{itemize}

\subsection{Filtering out of brane cosmological constant}

It is easy to see that brane induced gravity for $n>2$ is not very
sensitive to a brane cosmological constant: The addition of the brane
tension/cosmological constant term,  $S_\Lambda=-\Lambda\,\int
d^4\sigma\,\sqrt{-g(X(\sigma))}$, to the action amounts, at the
linearized level, to the shift    
\be 
\wt T_{\mu\nu}(k)\rightarrow \wt T\,'_{\mu\nu}(k)=\wt T_{\mu\nu}(k)
+\Lambda \,\eta_{\mu\nu}\,\delta^{(4)}(k)\,. 
\label{TM+Lambda}
\ee 
The validity of the perturbative analysis around flat background
requires that $\Lambda$ is small, of the same order as $T_{\mu\nu}$.
In ordinary Einstein-Hilbert gravity, $\Lambda$, however small,
destabilizes the flat space solution toward a de Sitter space (the 
manifestation of this at the propagator level is described below).
Hence, in a modified theory of gravity the stability of flat space in
the presence of $\Lambda$ is an indication that gravity has been made
less sensitive to the cosmological constant. The analogue of this
argument for Fierz-Pauli massive gravity is considered in \cite{DHK},
although the actual mechanisms differ. 

Let us write the solutions (\ref{s}) and (\ref{hp}) explicitly in
terms scalar and traceless-transverse components of $T_{\mu\nu}$,   
\be
\wt s(k)=-\frac{2}{3B}\,\frac{1}{k^2+\frac{A}{B}\,
\frac{d-2}{2(d-5)}\,\la\wt G\ra^{-1}}\,\wt T_s\,,\qquad
\wt h^\perp_{\mu\nu}=\frac{1}{B}\,\frac{1}{k^2-\frac{A}{B}\,
\la\wt G\ra^{-1}}\wt T^\perp_{\mu\nu} \,.
\label{s-hp-proj}
\ee 
As discussed in Appendix A, it is only in this form that the solutions
are valid for a cosmological constant source. For such a source, 
 $\wt T_{\Lambda\mu\nu}^\perp=0$ and $\wt T_{\Lambda s}=4\Lambda \,
\delta^{(4)}(k)$, leading to  $\wt h^\perp_{\Lambda\mu\nu}=0$ and,    
\be
\wt s_{\Lambda}(k)=4\,c\,\delta^{(4)}(k)\,,\qquad {\rm where}\qquad 
c=-\frac{4}{3}\,\frac{d-5}{d-2}\,\frac{\Lambda}{A}\, \la\wt G\ra(0)\,.
\label{s-hp-Lambda}
\ee 
The crucial point is that only for $n>2$, $\la\wt G\ra(0)$ is {\it
  finite} and the analysis well defined (for example, for a Gaussian
profile and for even $n$, $\la\wt G\ra(0)=-\,(\frac{n}{2}-2)!
/[(4\pi)^{\frac{n}{2}}\Gamma(\frac{n}{2})\omega^{n-2}]$). Thus, in
coordinate space, $\Lambda$ effectively shifts the background from
$s=0$ to a constant value $s=4c$, keeping the flat background
essentially unchanged. It is instructive to contrast this with
Einstein-Hilbert gravity where one would get $\wt s_{\Lambda}\sim
\Lambda\, \delta^{(4)}(k)/k^2$, which is a solution to $\Box_4
s_{\Lambda}\sim \Lambda$. Then $s_\Lambda$ is quadratic in distance
and diverges at large distances, indicating that flat space is not a
good starting point for a perturbative expansion of de Sitter space. In
contrast, in BIG, $\Lambda$ does not produce a non-flat metric on the
brane and is hence filtered out
\footnote{Codimensions $n=1,2$ give $\la\wt G\ra(0)=\infty$
  due to an IR divergent $q$-integral and the above discussion of
  filtering, formulated around flat background, does not apply. For
  the the $n=2$ case, see \cite{Corradini, Kaloper1,Kaloper2}.}. 

The mechanism by which BIG filters out $\Lambda$ is sometimes
heuristically explained on the basis of the behaviour of the $n=1$
model in which gravity becomes higher dimensional, and hence weaker,
at very large distances. But this model does not filter out $\Lambda$
and the $n>2$ model that does, has a different large distance
behaviour. So let us reiterate two relevant aspects of the above
analysis:
\begin{enumerate}
\item Only a non-zero $\la\wt G\ra(0)$ is relevant to filtering out
  $\Lambda$. The momentum dependence of $\la\wt G\ra$ (that
  contributes to graviton decay and a varying gravitational coupling)
  does not play any role in this, in contrast to the more general
  ``filter'' mechanisms of \cite{ADDG}
\item It also follows that at large distances ($k^2\rightarrow 0$),
  the behaviour of gravity does not become higher dimensional, unlike
  the $n=1$ DGP model. Hence the filtering of the cosmological
  constant cannot be attributed to such a behaviour.
\end{enumerate}
In BIG, the cosmological constant is completely filtered out
at the linearized level. One needs to go beyond the linear analysis to
make a statement about its remnant effects. In particular, explaining
the small observed value of $\Lambda$ (or dark energy density) is a
different issue not addressed by BIG at this stage.

\subsection{On the origin of the tachyon and ghost problems} 

As we have seen, a tachyonic ghost mode is encountered while analyzing
BIG around flat background \cite{DR}. Although flat space is a
valid solution to the $\Lambda=0$ model, one may still suspect that
the tachyon and ghost problems are related to this very set up. After
all, a generic $3$-brane always has a tension which, in some sense, is
the stuff that holds the brane together. So one always has to work
with the full action,
$$
S_{EH}^{bulk}+S_{EH}^{brane}+S^{brane}_\Lambda + S_m^{brane}\,.
$$ 
Even though a $\Lambda\rightarrow 0$ limit may exist, in that limit
$S_\Lambda^{brane}=\Lambda\int d^4\sigma \sqrt{-g}$ should be replaced
by the action for a tensionless brane \cite{Bo,Bozhilov} rather than
totally eliminated. But tensionless branes have special properties
that make them unsuitable for our purposes and hence need not be
considered. This line of argument tells us that the ``basic'' brane
induced gravity model reviewed in section 3 does not really
correspond to a brane setup, unless augmented by a brane tension
term. {\it Could this also provide a solution to the tachyon and ghost
  problems?} This possibility is discussed below. 

One approach to addressing this question is to consider, as suggested
in \cite{G-rev,DGS1,DGS2}, a large $S_\Lambda^{brane}$ so the matter
action $S_m^{brane}$ can be neglected to first approximation. This
leads to classical solutions for bulk gravity sourced by
$\Lambda$. Around this one should study metric perturbations sourced
by $S_m^{brane}$ and influenced by $S^{brane}_{EH}$. As pointed out
above, the existing solutions \cite{Gregory:1995qh,Charmousis} are not
adequate for this purpose since they do not satisfy the equations of
motion with the correct brane source. Hence, such analysis has not yet
been carried out (see footnote 3). 

However, the above parameter range ($S_\Lambda^{brane}>>S_m^{brane}$)
is not the only one to explore. In fact, we saw that the linearized
analysis is already capable of exploring the model in the regime where
$S_\Lambda^{brane}$ and $S_m^{brane}$ have comparable but small
contributions. Thus, it can also provide some insight into the
implication of the brane tension for the tachyon and ghost problems.

The tachyon and ghost problems showed up in the propagator (\ref{s})
for the $s$ field which has the general structure
$$
-\,\frac{2}{3}\,\frac{1}{Bk^2 -A\,\omega^{n-2}|N^{(0)}|\,\Sigma(k)}\,.
$$
The overall sign signals a ghost and the relative sign in the
denominator, a tachyon. The appearance of $A$ and $B$ coefficients
(associated with bulk and brane actions) shows that $\Sigma(k)$
(relevant to the tachyonic nature) arises entirely from integrating
out bulk modes, while the rest (the $-2/3$ factor and hence the ghost
nature) originates in the brane Einstein-Hilbert term. Let's emphasize
that the above form is a valid way of presenting and interpreting the
propagator as long as the $B$ term in the action dominates over the
$A$ term. This is the phenomenologically interesting case, with a very
light graviton and effectively $4$-dimensional gravity. It is also in
this range that the tachyon and ghost identifications
hold \footnote{For example, in the opposite case when the $A$ term
  dominates over the $B$ term (as in string theory setups), the
  propagation is described by the $\Sigma(k)$ term and gravity is
  healthy and essentially $d$-dimensional.}.

\begin{itemize}
\item {\bf The tachyon problem:} Now, this is not hard to dismiss. In
  the previous subsection we saw that the ``tachyonic'' mass of $s$
  (more precisely, the $\Sigma(0)\neq 0$ part) was instrumental in
  ``absorbing'' $\Lambda$ and keeping the brane metric flat. On the
  other hand, the classical solutions of
  \cite{Gregory:1995qh,Charmousis} indicate that $\Lambda$ can equally
  well be absorbed by curving directions transverse to the brane,
  keeping the brane metric flat. The comparison indicates that
  approximating the bulk as flat, has to force a mass on $s$ in order
  for the brane metric to remain flat in the presence of $\Lambda$. In
  this sense, {\it the mass is at least partly an artifact of the flat
    background approximation and is functionally equivalent to a
    ``healthy'' bulk curvature to be sourced by $\Lambda$}. This
  demonstrates that the ``tachyonic'' mass is not an unhealthy
  feature. Indeed, $m^2_s$ is tachyonic only with respect to the
  ghost-like kinetic term of $s$, and as such has a positive
  contribution to the $s$ field Hamiltonian $\sim \int(-\dot s^2 -
  \nabla s^2+m^2_s s^2)$.

\item {\bf The ghost problem:} This however is more fundamental and
  may not be cured even if the flat background is replaced by a curved
  one sourced by the brane tension . The reason is that even in such a
  curved background, the brane metric remains flat and hence its
  fluctuations are described by $S^{brane}_{EH}$, expanded around flat
  4-dimensional spacetime. This contributes a factor $-(2/3)(1/Bk^2)$
  to the scalar mode propagator, exactly as in flat background. A
  modified ``self-energy'' term, $\Sigma_{curved}$, arises from
  integrating out bulk modes, but now it also depends on the brane
  tension. As before, small graviton masses and effectively
  $4$-dimensional gravity require a large $B$ coefficient. Thus the
  overall propagator again has the standard $4$-dimensional tensor
  structure, corrected by self-energy terms. Now, it is well known
  that a propagator with a tensor structure associated with the
  $4$-dimensional Einstein-Hilbert action is ghost free only in the
  massless limit (where the scalar mode ghost cancels against a
  contribution from the spin $2$ mode). As soon as the mass poles are
  shifted away from zero, the scalar mode ghost remains uncanceled.
\end{itemize}

An implication is that the ghost can be avoided only if
$\Sigma_{curved}(0)=0$. In a flat space background, this would have
prevented the theory from filtering out the cosmological constant. But
in a curved background, where $\Lambda$ is already absorbed in the
background curvature, this could lead to a consistent theory of
massless gravity with filtered out $\Lambda$. However, it is not
obvious that a background sourced by the brane tension will lead to
such a self-energy contribution. In fact the opposite seems to be the
case by a continuity argument: In the small $\Lambda$ limit, the
expressions should reduce to what we have already computed, in which
case $\Sigma(0) \neq 0$. Hence we conclude that {\it in the
  phenomenologically interesting parameter ranges, the scalar ghost
  will be a generic feature of brane induced gravity in any
  background, except those for which $\Sigma_{curved}(0)=0$}.

The absence of a ghost in the $d=5$ DGP model is a numeric
coincidence. On general grounds, reduction of gravity from $d$ to $4$
dimensions gives a factor $d-5$ in the scalar sector. In the BIG
setup, this results in solutions with $s=0$, thus evading the ghost
problem.

\subsection{Implication for couplings}
The structure of the solution (\ref{s-hp-Lambda}) essentially 
corresponds to a warped metric with its standard implications for the
couplings. The metric in the presence of $\Lambda$ and matter sources
becomes,   
\be
g_{\mu\nu}(x)=(1+c)\,\eta_{\mu\nu} +  h_{\mu\nu}^{(m)}(x)\,,
\ee
where $c\sim \Lambda/(A\,\omega^{n-2})>0$ and $h_{\mu\nu}^{(m)}$
is the metric perturbation sourced by matter. Adopting the standard
convention to use $\eta_{MN}$ as the flat spacetime metric, one
defines the physical metric      
\be
g'_{\mu\nu}=\frac{g_{\mu\nu}}{(1+c)}\,.
\ee
In terms of this, $B\int{\sqrt g}R=(1+c)B\int {\sqrt g'}R'$. For the
matter action, $\int{\sqrt g}{\cal L}(g,\psi,\lambda)=
(1+c)^2\int{\sqrt g'}{\cal L}(g',\psi',\lambda')$, where the matter
fields, collectively denoted by $\psi$, and their couplings $\lambda$,
have to be scaled appropriately for the equivalence principle to hold.
This results in an effective $B'=B/(1+c)$ or an effective Newton
constant $G_N'=(1+c)G_N$, consistent with related observations
in \cite{DGHS,DGS1,DGS2,G-rev,Kaloper1,Kaloper2}. Although this
analysis is valid for small $c$ (so that our perturbative treatment
remains valid) it indicates that for a large $\Lambda$ one needs a
starting $B$ larger than the observed $M_p^2$ (for $c>0$). This can
relax the constraint on the bulk coupling $A$ based on the smallness of
the graviton mass by a factor $(1+c)$ which is welcome news. However,
it may also introduce new hierarchy issues for other standard model
couplings. 

\subsection{Contrast with Fierz-Pauli Massive gravity}

In the ghost-free FP massive gravity theory around flat background,
reviewed in a later section, the spin-2 graviton $h^\perp$ has a mass
$m_2$, whereas, to avoid the ghost, the mass $m_0$ of the scalar mode
$s$ is sent to infinity. Hence for any matter source $T_{\mu\nu}$,
$s=0$. The caveat is that a constraint equation one gets for $s$ (for
$m_0=\infty$) only implies $\Box_4 s=0$ so, in general,
$s=c_1+c_2s^{harmonic}$. Substituting back into the equations, gives
$c_2=0$ for any source and $c_1=-4\Lambda G_N/3m^2_2\neq 0$ only for a
cosmological constant source. To contrast the two theories, in BIG,
$s$ is always sourced and remains well behaved in the presence of
$\Lambda$ due to its own mass, but in the ghost free FP gravity, $s$
is zero except for a $\Lambda$ source and the screening parameter is
the mass of spin-2 graviton $m_2$, which is itself not sourced by
$\Lambda$! A FP theory with finite $m_0$ (and hence a ghost), when
sourced by $\Lambda$, would behave similar to BIG for $n>2$, with
$m_0$ as the screening parameter. However, the $m_0\rightarrow \infty$
limit is not continuous. Thus, although superficially similar, BIG
with $n>2$ and FP massive gravity use different mechanisms to filter
out $\Lambda$.

\section{Inclusion of Extrinsic Curvature Terms}

We now solve the modified equations of motion after the inclusion of
the extrinsic curvature terms. These are ubiquitous in any brane
induced gravity setup and arise at the same perturbative order as the
induced Einstein-Hilbert term \cite{Corley, Ardalan, Cheung}.  The
brane transverse fluctuations enter as new degrees of freedom. The
brane tension can be included in $T_{\mu\nu}$.  Again the analysis is
performed in terms of gauge invariant variables. It turns out that the
extrinsic curvature terms have {\it no effect whatsoever} on the
solutions for the brane fields. While this does not cure the
tachyon/ghost problem, it at least insures that $4$-dimensional
gravity is not modified in other unwanted ways by the extrinsic
curvature terms. The negative result is essentially due to the $Z_2$
reflection symmetry of the background about the brane
position. Breaking this symmetry will lead to extra non-trivial
contributions from the extrinsic curvature terms. With this summary,
the reader not interested in the details of the calculation can safely
skip this section.
 
\subsection{Extrinsic curvature term in the action} 
The contribution of the extrinsic curvature to the action is
\cite{Corley} 
\be
S_\Omega=C\int d^4\sigma \sqrt{-g}\,\,\left(\Omega^{M}_{\alpha\beta}\,\, 
\Omega_{M}^{\,\,\,\,\alpha\beta}-\Omega^{M \alpha}_{\alpha}\,\,
\Omega_{M \beta}^{\,\,\,\,\,\,\,\,\,\beta} 
\right)\,,
\ee
where
\be
\Omega^M_{\alpha\beta}=\p_\alpha\p_\beta x^M
-\gamma^\lambda_{\alpha\beta}\,\p_\lambda x^M + \Gamma^M_{NK}\,
\p_\alpha x^N\p_ \beta x^K\,.
\ee
Here $\gamma$ and $\Gamma$ are the Christoffel connections
corresponding to the metrics $g_{\alpha\beta}(x(\sigma))$ and
$G_{MN}(x(\sigma))$, respectively. For a thin brane the quadratic
action for the fluctuations can be worked out by setting
$g=\eta+h$,  $G=\eta+H$ and $\delta X^M=\{f^\mu, y^i\}$,
\bea
S_\Omega&=&C\int dx_{||}^4\Big[\half\left(\p_\alpha H_{i\beta}
\p^\alpha H^{i\beta} -\p_\alpha H^{\alpha i}\p^\beta H_{\beta i}\right)
+\p_\alpha H^{\alpha i}\p_i H^\beta_\beta - \p_\alpha H^{i\beta}
\p_i H^{\alpha\beta}\nn \\
&&\qquad+\frac{1}{4}\left(\p_i H_{\alpha\beta}\p^i H^{\alpha\beta} -
\p_i H^\alpha_\alpha\p^i H^\beta_\beta\right)+y^i\p_i\left(
\Box_4 H^\alpha_\alpha-\p_\alpha\p_\beta H^{\alpha\beta}\right) 
+\cdots\Big]\,, 
\eea
where the ellipsis stand for functions of $h_{\mu\nu}-H_{\mu\nu}$ that 
do not contribute to the equation of motion for a metric variation
$\delta h_{\mu\nu}=\delta H_{\mu\nu}$.

As described in section 2, for thick branes, $H_{MN}(x(\sigma))$ are
replaced by $\la H_{MN}\ra (x_{||})$. A new feature of $S_\Omega$ is
that it contains transverse derivatives of $H$ restricted to the
brane, $\p_i H|_{brane}\equiv (\frac{\p}{\p x^i_\perp} H)(x_{||},
x^i_\perp=y^i_0)$. For a thick brane, these should be replaced by
$\la\p_i H \ra\equiv \la\frac{\p}{\p x^i_\perp} H\ra (x_{||})$.
The $y^i(x_{||})$ are brane fields and do not need any
``blurring''. Hence, for thick branes, the appropriate action is,   
\bea
S_\Omega&=&C\int dx^4\Big[\half\left(\p_\alpha \H_{i\beta}
\p^\alpha\H^{i\beta} -\p_\alpha\H^{\alpha i}\p^\beta\H_{\beta i}\right)
+\p_\alpha\H^{\alpha i}\la\p_i H\ra^\beta_\beta \nn\\
&&\qquad -\p_\alpha\H^{i\beta}\la\p_i H\ra^{\alpha\beta}
+\frac{1}{4}\left(\la\p_i H\ra_{\alpha\beta}\la\p^i H\ra^{\alpha\beta}
-\la\p_i H\ra^\alpha_\alpha\la\p^i H\ra^\beta_\beta\right)\nn \\
&&\qquad +y^i \left(\Box_4\la\p_i H\ra^\alpha_\alpha-\p_\alpha\p_\beta 
\la\p_iH\ra^{\alpha\beta}\right)+\cdots\Big]\,.
\label{blSOmeg}
\eea

\subsection{Equations of motion with extrinsic curvature contributions} 

To obtain the contribution of $S_\Omega$ to the equations of motion
for a thick brane, one considers variations $\delta y^i$, $\delta\H$,
$\delta\la\p_i H\ra$ and re-expresses the last two in terms of
the bulk variation $\delta H$. In particular,   
\bea
\delta\la\p_i H\ra_{MN}&=&\int d^nx_\perp P(x_\perp-y_0)\,
\p_i\delta H_{MN}(x_{||},x_\perp-y_0) \nn\\
&=&-\int d^nx_\perp\p_i P(x_\perp-y_0)\,\delta H_{MN}(x_{||},x_\perp-y_0)\,. 
\eea
Then for $\delta H$ the variation can be written in a compact from as,  
\bea
\delta S_\Omega=-2C\int d^dx\left[P(x_\perp-y_0)\,\p_\alpha\,
{\cal N}^{i\alpha\beta}\,\delta H_{i\beta} -\half \p_i P(x_\perp-y_0)\,
{\cal N}^{i\alpha\beta}\,\delta H_{\alpha\beta}  \right]\,,
\eea
where, to linear order, 
\bea
{\cal N}^i_{\alpha\beta}(x_{||}) &=&
\la\Omega^i_{\alpha\beta}-\Omega^{i\,\,\lambda}_{\lambda}\eta_{\alpha\beta}\ra  
= \p_\alpha\p_\beta y^i +\half\left(\p_\alpha\H^i_\beta +
\p_\beta\H^i_\alpha-\la\p^iH\ra_{\alpha\beta}\right) \nn\\
&&\hspace{4cm}-\left(\Box_4 y^i+\p_\lambda\H^{\lambda i}-\half\la\p^i
H\ra^\lambda_\lambda\right)\,\eta_{\alpha\beta} \nn 
\\
&=&\p_\alpha\p_\beta F^i+\half\left(
\p_\alpha\Hp^i_\beta+\p_\beta\Hp^i_\alpha-\la\p^iH^\perp\ra_{\alpha\beta}
\right) \nn\\
&&\hspace{1cm}-\left(\Box_4 F^i+\p_\lambda\Hp^{\lambda i}-\half\la\p^i
H^\perp\ra^\lambda_\lambda\right)\,\eta_{\alpha\beta} 
+\frac{3}{2d}\eta_{\alpha\beta}\la\p^i S\ra\,. \nn
\eea
$F^i$ are gauge invariant extensions of $y^i$ (the analogues of  
$F^\mu$ in transverse directions), 
\be
F^i= y^i + \la A^i\ra +\half \la\p^i\Phi\ra \,.
\label{Fi}
\ee
Now, from $\delta(S^{bulk}_{EH}+S^{brane}_{EH}+S^{brane}_\Omega)$ one
obtains the thick brane metric equation of motion with extrinsic
curvature contributions contained in the $C$-dependent terms, 
\bea
&&\frac{A}{2}\left[\Box_d H^{\perp MN}+\frac{d-2}{d}
\left(\p^M\p^N-\eta^{MN}\Box_d\right)S \right]\nn\\
&&+ P(x_\perp-y_0) \Bigg[\frac{B}{2}\Big(\Box_4 h^{\perp \mu\nu} 
+\frac{1}{2}\left(\p^\mu\p^\nu-\eta^{\mu\nu}\Box_4\right)
s \Big)\delta^M_\mu\delta^N_\nu -C\,\p^\lambda{\cal N}^i_{\lambda\nu}\,
(\delta^M_i\delta^N_\nu+\delta^M_\nu\delta^N_i)\Bigg] \nn\\
&&\hspace{2.5cm} +\,C\,\p_iP(x_\perp-y_0)\,{\cal N}^{i\mu\nu}\,
\delta^M_\mu\delta^N_\nu=-\frac{1}{2}P(x_\perp-y_0)\,T^{\mu\nu}\,
\delta^M_\mu\delta^N_\nu\,. 
\label{eomExt}
\eea

Also, in the quadratic action $S_\Omega$, the $y^i$ appear as Lagrange
multipliers and their equation of motion is the constraint equation,
\be
\p^\mu\p^\nu{\cal N}^i_{\mu\nu}\equiv 
\half\la\p^i\left(\Box_4 H^{\perp\lambda}_\lambda-\p^\mu\p^\nu
H^\perp_{\mu\nu}+\frac{3}{d}\Box_4 S\right)\ra =0\,. 
\label{eomy}
\ee
But we will see that this is already contained in (\ref{eomExt}). The
expression within braces is the scalar curvature of the metric
$(\eta+H)_{\mu\nu}$. 

\subsection{Solutions with extrinsic curvature contributions}

Below we show that the inclusion of $C$-dependent terms in
(\ref{eomExt}) has no effect on the solutions for brane fields. We
closely follow the steps in the $C=0$ case.    

{\it The $\eta_{MN}$-trace} of (\ref{eomExt}) gives $S$ in terms of
$s$ and ${\cal N}^i_{\mu\nu}$. On transforming to momentum space,
one gets the solution 
\be
\wt S(k,q)=\frac{-d}{A(d-1)(d-2)}\,\,\frac{\wt P(q)}{k^2+q^2}
\,\,\left(\wt T^\mu_\mu+\frac{3}{2}B\, k^2\, \wt s +i2C\,q_i\, 
\wt{\cal N}^{i\,\,\lambda}_\lambda \right)\,.
\label{SExt}
\ee
The relevant feature of this is that, since $\wt{\cal
  N}^{i\,\,\lambda}_\lambda$ depends only on $k$, the $C$-dependent
term is an odd function of $q^i$ and $\la q_i\,\wt{\cal N}^{i\,\,
\lambda}_\lambda\ra=0$. Therefore, it does not contribute to
$\la\wt S\ra$ which is, then, still given by (\ref{blS}). For later
convenience, we record the analogue of (\ref{SbrakS}), 
\be
\wt S(k,q)=-\frac{\wt P(q)}{k^2+q^2}\,\,\left[\frac{\la\wt  S\ra}
{\la\wt G\ra}\,\,+i\,\frac{2d}{A(d-1)(d-2)}\,C\,q_i\,
\wt{\cal N}^{i\,\,\lambda}_\lambda\right]\,.
\label{SbrakSN}
\ee

{\it The $(i,j)$ Components} of (\ref{eomExt}) are not affected by the
$C$ dependent terms. Hence they yield again the solution (\ref{HijS})  
for $H^\perp_{ij}$ in terms of $S$. The solution (\ref{SbrakSN}) for
$S$ then gives the same expressions for $\la\wt H^{\perp j}_j\ra$ and
$\la q^i q^j\wt H^{\perp ij}\ra$ as in (\ref{trHqqH}). 

Manipulations of the surface equation also lead to the same result as
the $C=0$ case. Hence $\la S\ra$ and $s$ are still related by
(\ref{Ss}) and finally the solution for $s$ is given by (\ref{s}). 
 
{\it The $(\mu,i)$ Components} of (\ref{eomExt}) contain a
$C$-dependent term and give,
\be
\wt H^{\perp\mu i}(k,q)=-\frac{d-2}{d}\,\,\frac{k^\mu q^i}{k^2+q^2}
\,\,\wt S(k,q)-2\,i\,\frac{C}{A}\frac{\wt P}{k^2+q^2}k^\lambda
\wt {\cal N}^{i\mu}_\lambda \,.
\label{HmuiExt}
\ee
However, $\la q_i\wt H^{\perp\mu i}\ra$ is still given by (\ref{qvecH}).
Hence going through the same steps as before, we get the same solution
(\ref{Fmu}) for $F_\mu$. 

{\it The $(\mu,\nu)$ Components} of the equation of motion in 
momentum space is,  
\bea
&&\hspace{-1.2cm}\frac{A}{2}\left[(k^2+q^2)\wt
  H^\perp_{\mu\nu}+\frac{d-2} {2}\left(k_\mu
  k_\nu-\eta_{\mu\nu}(k^2+q^2)\right)\wt S\right]\nn\\ 
&&+\wt P(q)\left[\frac{B}{2}\left( k^2
  \wt h^\perp_{\mu\nu}+\half\left(k_\mu k_\nu-k^2\eta_{\mu\nu}\right)
  \wt s\right)-iC q_i\,\wt{\cal N}^i_{\mu\nu}(k) -\frac{1}{2}
  \wt T_{\mu\nu} \right]=0 \,.
\label{munuExt}
\eea 
When we solve this for $H^\perp_{\mu\nu}$ and evaluate $\la\wt
H^\perp\ra_{\mu\nu}$, the $C$-dependent term drops out (because of
integration over an odd function of $q^i$). Using this to compute    
$\wt h^\perp_{\mu\nu}$ again gives the same result (\ref{hp}) as the
$C=0$ case.  

{\it The $y^i$ Equation of motion} (\ref{eomy}) follows also as a
consequence of the metric equation: contract (\ref{munuExt}) with
$k^\mu k^\nu$, use $k^\mu k^\nu \wt H^\perp_{\mu\nu}=q^i q^j \wt
H^\perp_{ij}$ and (\ref{HijS}) to eliminate $\wt H^\perp_{ij}$ in
favour of $\wt S$. This gives (\ref{eomy}) in momentum space. 

Thus, this constraint does not affect the solutions for the brane
fields, but it provides an equation for $F^i$ as follows: Using
$H^{\perp\lambda}_\lambda=-H^{\perp i}_i$, $\p^\mu\p^\nu 
H^\perp_{\mu\nu} =\p^i\p^j H^\perp_{ij}$ and (\ref{HijS}) to rewrite 
these in terms of $S$, one gets
\be
k^\mu k^\nu\, \wt {\cal N}^i_{\mu\nu}= -\frac{i}{2d}(d-1)(d-5)\,k^2
\,\la q^i\,\wt S\ra =0 \,.
\label{NS}
\ee
Ignoring the 4-dimensional harmonic solutions, this gives 
$\la q^i\, \wt S\ra=0$, or $\la\p^i\,S\ra =0$.
Using the solution for $S$ in (\ref{SExt}), this becomes (with no sum
over $j$), 
\be
\la q^j\wt S\ra= -i\,
\frac{2d\,C}{A(d-1)(d-2)}
\,\,\left[\int d^nq\,\frac{\wt
    P^2(q)(q^j)^2}{k^2+q^2}\right] 
\,\,\left(\,\wt{\cal N}^{j\,\,\lambda}_\lambda\right)
=0\,,
\ee 
which holds only if,  
\be
\wt{\cal N}^{j\lambda}_{\lambda}(k)\equiv 3k^2\wt F^j+3i\la q_i
\wt H^{\perp ji}\ra-i\frac{3}{2}\la q^j\wt H^{\perp i}_i\ra + i
\frac{6}{d}\la q^j \wt S\ra =0\,.
\ee
Again, expressing $H^\perp_{ij}$ in terms of $S$ gives
$$
3 k^2\left(\wt F^j(k)+i\frac{d-2}{2d}\,\la\frac{q^j\wt S}{k^2+q^2}\ra 
\right)=0 \,.
$$
But (\ref{SbrakSN}) implies $\la q^j\wt S/(k^2+q^2)\ra \propto 
\wt{\cal N}^{j\lambda}_{\lambda}=0$, so we finally get the $F^i$
equations, 
\be
\Box_4\, F^j=0\,.
\ee
Hence, up to harmonic functions, $F^i=0$. This is a consequence of the
reflection symmetry of the set up in directions transverse to the
brane. A background that breaks this symmetry can lead to 
non-trivial $F^i$. Note that for the DGP model with $d=5$, the
constraint equation (\ref{NS}) is trivial and does not imply an
equation for $F^5$.   
 
Thus the presence of the extrinsic curvature related corrections has
no effect on BIG at the linearized level, in particular not on the
ghost problem. While this is discouraging, the positive aspect is
that such terms (that on general grounds are ubiquitous
in all braneworld models) do not cause large deviations from standard 
gravity.

\section{Revisiting Fierz-Pauli Massive Gravity}

In the remaining part of this paper, we turn to a comparison of brane
induced gravity for $n>2$, with massive versions of Einstein-Hilbert
gravity in $4$ dimensions. In this section we start with a review
Fierz-Pauli massive gravity theory \cite{FP1,FP2} (for reviews, see
\cite{RubTin,Bebronne}). The theory is then rewritten in a way that
facilitates comparison with the $4$-dimensional effective action for
BIG, to be derived in the next section. {\it In fact, the construction
  presented here for the FP theory enables us to make sense of the BIG
  effective action}. 

\subsection{Fierz-Pauli massive gravity in gauge invariant variables}

In standard 4-dimensional Einstein-Hilbert gravity, metric
fluctuations $h_{\mu\nu}$ in a flat background can be made massive by
introducing mass parameters, say, $a$ and $b$. The combination $a+b=0$
results in the ghost free Fierz-Pauli massive gravity theory
\cite{FP1,FP2}. Since we are interested in the origin of ghost in
brane induced gravity, we keep $a$ and $b$ unconstrained and use the
term ``Fierz-Pauli theory'' in this generalized sense.

The gauge invariance of linearized gravity, broken by the mass terms,
is restored by the St\"uckelberg technique. This amounts to performing
a gauge transformation and retaining the gauge parameters as new
fields. The resulting action is,      
\be
S_{FP}=S_{EH}[h] -\frac{B}{4} \int d^4x \left[
a(h_{\mu\nu}+2\p_{(\mu}\bar f_{\nu)})^2
+b (h^\mu_\mu+2\p_\mu \bar f^\mu)^2\,
\right] \,.
\label{S-FP}
\ee
$S_{EH}[h]$ is read off from (\ref{S-bulk}) for $d=4$. The $\bar
f_\mu$ are St\"uckelberg fields transforming as $\delta\bar f_\mu
=\lambda^\mu$. Combined with $\delta h_{\mu\nu}=-2\p_{(\mu}
\lambda_{\nu)}$, it keeps $S_{FP}$ invariant. The $\bar f_\mu$
can be interpreted as the Goldstone fields corresponding to the
broken symmetry \cite{AGS}. This conveys a sense of graviton  masses
arising due to spontaneous symmetry breaking, even in the absence of a 
more detailed fundamental mechanism.

The theory can be rewritten in terms of $h^\perp$ and $s$ of
(\ref{h-giv}) and the new gauge invariant variables,   
\be
\bF_\mu=\bar f_\mu+a_\mu+\frac{1}{2}\p_\mu\phi \,.
\label{F-FP}
\ee
Note that these are the same as the $F_\mu$ in (\ref{F}) once we
identify the St\"uckelberg field as $\bar f_\mu = f_\mu+\la 
A_\mu\ra+\frac{1}{2}\p_\mu(\la\Phi\ra)$, hence the similar 
notation. The Fierz-Pauli action (\ref{S-FP}) in terms of the gauge
invariant variables becomes,
\bea
S_{FP} &=& -\frac{B}{4}\int d^4x \Big[
h^{\perp\mu\nu}(\,-\Box+a\,)h^\perp_{\mu\nu}
+s(\,\frac{3}{8}\Box +\frac{a}{4}+b\,)s
+4(\frac{a}{4}+b)\p\cdot\bF \,s \nn\\
&&\qquad\qquad\qquad +\, a\,\left(\p_\mu\bF_\nu-\p_\nu\bF_\mu)^2  
+4\,(a+b)\,(\p\cdot\bF)^2 \right) \Big]\,.
\label{SFP}
\eea 
The equation of motion for $\bF_\mu$,
\be
a\,\Box\bF_\mu+(a+2b)\p_\mu(\p\cdot\bF)+(\frac{a}{4}+b)\p_\mu s=0\,,
\label{eomF-FP}
\ee
has the purely longitudinal solution,
\be
\bF_\mu=-\frac{\frac{a}{4}+b}{2(a+b)}\,\Box^{-1}\p_\mu s \,.
\label{F-FPsoln}
\ee
Then, the $h_{\mu\nu}$ equation (obtained after expressing variations 
$\delta h^\perp$ and $\delta s$ in terms of $\delta h$ using the
projection operators in Appendix A),
\be
(-\Box+a)h^\perp_{\mu\nu}+\frac{4}{3}(\eta_{\mu\nu}-
\frac{\p_\mu\p_\nu}{\Box})\left((\,\frac{3}{8}\,\Box +\frac{a}{4}+b\,)s
+2(\frac{a}{4}+b)\p\cdot\bF\right)=T_{\mu\nu}/B\,,
\label{eomh-FP}
\ee
has the massive graviton solutions,
\be
\wt h^\perp_{\mu\nu}=\frac{1}{B}\frac{1}{k^2+a}\,\wt T_{\mu\nu}^\perp
\,,\qquad \wt s=-\frac{2}{3B}\frac{\wt T}{k^2-\frac{1}{2}
\frac{a}{a+b} (a+4b)}\,,
\label{hsFP}
\ee
corresponding to masses $m_{h^\perp}^2=a$ and $m_s^2=-\frac{1}{2}
\frac{a}{a+b}(a+4b)$. The full metric is obviously, 
$$
h_{\mu\nu}=h_{\mu\nu}^\perp+\frac{1}{4}\eta_{\mu\nu}s+\p_\mu\bF_\nu+
\p_\nu\bF_\mu \,,
$$
with the gauge choice $\bar f_\mu=0$. In the following paragraphs, we 
comment on the ghost free theory before returning to the general
case in the next subsection. 

The massive FP gravity theory is, in general, not ghost free: A
standard analysis of the gauge invariant amplitude (\ref{gia}) for
interaction between sources $T_{\mu\nu}$ and $T'_{\mu\nu}$ shows that
for $a\neq 0$, the $s$ field is always a ghost\footnote{The presence
  of the ghost for $a+b\neq 0$ can already be understood at the level
  of action (\ref{SFP}) in terms of the St\"uckelberg fields: The
  vector field $\bF_\mu=F_\mu^\perp+\p_\mu\phi_F$ has an unhealthy
  kinetic term because $(a+b)(\p_\mu\bF^\mu)^2$ leads to a
  $4$-derivative term for the longitudinal mode $\phi_F$, resulting in
  ghost non-decoupling. For $a+b=0$, the $4$-derivative term is
  eliminated. This provides an intuitive way of detecting the massive
  gravity ghost.}, as shown, for example, in \cite{FP1,vDVZ1,VN} or
more explicitly in \cite{Nunes}. The only way to get rid of the ghost
is to set $a+b=0$ which gives $m_s=\infty$ and $s=0$. The outcome is
the ghost-free Fierz-Pauli massive gravity with\footnote{The origin of
  vDVZ \cite{vDVZ1,vDVZ2,vDVZ3} discontinuity, that arises in the
  massless limit of ghost-free FP theory, is manifest in this
  formulation: Sending $m_{h^\perp}\rightarrow 0$, keeping
  $m_s=\infty$ fixed, does not recover the standard massless solutions
  (\ref{hsd=4}) as $s$ remains zero and the theory does not couple to
  $T^\mu_\mu$.  The difference between the two massless theories does
  not affect light for which $T^\mu_\mu=0$. But matter is affected
  differently and the definitions of the Newton constant in the two
  theories differ by a factor of $3/4$. This shows up in comparing the
  bending of light in the two cases. The discontinuity is avoided in
  the non-linear theory through the Vainshtein effect
  \cite{Vainshtein,Deffayet}.},
$$
\qquad\wt h_{\mu\nu}=\frac{1}{B}\left(\frac{1}{k^2+a}\,\wt T_{\mu\nu}^\perp
+\frac{1}{3}\,\frac{k_\mu k_\nu}{a\,k^2}\,T \right)\,,
\quad (a+b=0)\,.
$$
An alternative solution to the ghost problem is to regard gravity as
an effective theory and give the ghost a mass above the scale of
validity of the effective theory by setting $a+b=\delta$ where
$\delta$ is positive and very small. Then the inaccessible ghost pole
becomes physically irrelevant \cite{Dvali:2008em}.

Let us discuss an issue raised in section 4.4. While the ghost free
theory has $s=0$ for a generic source, there is a caveat when
$T_{\mu\nu}$ corresponds to a cosmological constant: When $a+b=0$,
(\ref{eomF-FP}) implies $\Box_4 s=0$ and its solution $s=c_1+c_2
s^{harmonic}$ replaces the expression for $s$ in terms of $\p\cdot F$
obtained from (\ref{F-FPsoln}). In general $c_{1,2}=0$ by boundary
conditions, but for a cosmological constant source, the $h_{\mu\nu}$
equation gives $s=c_1=-4\Lambda/3aB$. In section $4.4$, this behaviour
was contrasted to that of brane induced gravity. For generic $a$ and
$b$ the two theories respond to the cosmological constant in the same
way.

\subsection{A $1$-parameter family of actions for FP gravity}

From the outset it was clear that massive gravitons and ghosts were
common features in both Fierz-Pauli theory (for $a+b\neq 0$) and BIG
(for $n>2$) at the linearized level. Since FP gravity is simpler, it
is natural to use it as a prototype to model some aspects of BIG, as
in \cite{DHK,Dvali:2008em}. But how much can the similarities be
stretched? To quantify this, below we rewrite the Fierz-Pauli action
in a way that helps us compare with, and make sense of, the BIG
$4$-dimensional effective action, to be derived in the next section.

As far as classical solutions are concerned, the action (\ref{SFP}) is
not unique. To find other equivalent actions, use the relation between
$s$ and $\p\cdot\bF$ obtained from (\ref{F-FPsoln}) to convert a part
of the $s^2$ term into $(\p\cdot\bF)^2$ and $s\p\cdot\bF$ terms. This
gives, 
\bea
S_{FP} &=& -\frac{B}{4}\int d^4x \Big[
h^{\perp\mu\nu}(-\Box +a)h^\perp_{\mu\nu}
+\frac{3}{8}s\Box s
+\left(\frac{a}{4}+b+\bw_1\right)ss \nn\\
&&+\, 4(\frac{a}{4}+b+\bw_2)\p\cdot\bF \,s 
+ a\,\left(\p_\mu\bF_\nu-\p_\nu\bF_\mu)^2  
+4\,(a+b+\bw_3)\,(\p_\mu\bF^\mu)^2 \right) \Big]\,,
\label{SFPw}
\eea 
where, $\bw_1 s^2+4 \bw_2 \p\cdot\bF s +4 \bw_3 (\p\cdot\bF)^2=0$
should hold on the classical solution. The $\bw$'s are further
constrained by requiring that the above action leads to the correct
equations of motion for $\bF_\mu$ and $h_{\mu\nu}$. The outcome is, 
\be
\bw_3 =\frac{a+b}{a/{4}+b}\, \bw_2 \,,\qquad 
\bw_2=\frac{a+b}{{a}/{4}+b}\, \bw_1 \,.
\ee
Thus we get a $1$-parameter family of actions, in terms of gauge
invariant variables, producing the same equation of motion as the 
Fierz-Pauli theory. 

A point to note is that, except for $\bw_1=0$, these equivalent
actions will look non-local when expressed in terms of $h_{\mu\nu}$,
with $\Box^{-1}$-type non-localities arising from the projection
operators (\ref{PO}) in Appendix A.

The particular choice $\bw_3=-(a+b)$ (implying $\bw_2=-(a/4+b)$)
decouples gravity from $\bF_\mu$ which become a free Abelian gauge
field. The resulting action involves only $h^\perp$ and $s$
(corresponding to integrating out $\bF_\mu$), 
\bea
S_{FP}=-\frac{B}{4}\int d^4x\Big[h^{\perp\mu\nu}(-\Box+a)
h^\perp_{\mu\nu}+\frac{3}{8}s\left(\Box+\frac{a(a+4b)}{2(a+b)}\right)s 
\Big]\,.
\label{SFPnoF}
\eea 
Note that, although the solutions (\ref{hsFP}) can be readily obtained
from here, this choice of $\bw_3$ is a singular limit of (\ref{SFPw})
in the sense that $\p\cdot F$ becomes undetermined \footnote{This is
  obvious from the $F_\mu$ equation of motion for (\ref{SFPw}) which
  gives, $2(a+b+\bw_3)\p\cdot F=-(a/4+b+\bw_2)s$}. 

When re-expressed in terms of $h_{\mu\nu}$, using projection operators
of Appendix A, this action remains manifestly gauge invariant, with no
dependence on St\"uckelberg fields and no sign of symmetry breaking.
But, as a trade off, it will contain $\Box^{-1}$-type non-localities
arising from projection operators. In this example, we know that these
non-localities arise as a result of integrating out the $\bF_\mu$ and
indicate the existence of a local formulation ((\ref{S-FP} or
(\ref{SFP})) in which St\"uckelberg fields are needed to restore gauge
invariance. The two cases are connected by the interpolating action
(\ref{SFPw}). {\it This construction is useful for obtaining the $4$-d
  effective action in brane induced gravity where the above steps are
  traversed in the reverse order}.

\section{$4$-d Effective Action in BIG and Broken Gauge Invariance}

In order to better understand the $4$-dimensional structure of BIG, we
need an effective $4$-dimensional description of it. In this section
we work out such an effective action by integrating out parts of the
fields with support away form the brane as well as the extra
components of the metric. Then we construct an action containing the
St\"uckelberg-like fields $F_\mu$, based on our understanding of
massive gravity in the previous section. 

This has implications for the origin of ghost and for the realization
of the brane general covariance as a spontaneously broken gauge
symmetry: The $F_\mu$ or $f_\mu$ (\ref{F}) of BIG are closely related
to St\"uckelberg fields $\bar f_\mu$ in Fierz-Pauli theory (more
precisely, $\bar f_\mu \sim f_\mu+\la A_\mu\ra+\frac{1}{2}
\p_\mu(\la\Phi\ra)$). Now, $\bar f_\mu$ restore gauge invariances
broken by the FP mass terms and, hence, can be interpreted as the
Goldstone fields associated with the broken symmetry \cite{AGS}. The
appearance of their analogues $f_\mu$ in BIG is thus a
manifestation of spontaneous breakdown of $4$-dimensional general
covariance by the bulk-brane setup. In this sense, BIG provides a
realization of the gravitational Higgs mechanism.

\subsection{$4\,$-dimensional effective action in Brane Induced Gravity}

In this subsection we derive a $4$-dimensional effective action for
BIG by integrating out all bulk related modes in the $A$ and $B$ terms
of (\ref{Sbig-I}). The bulk field $H_{MN}(x_{||}, x_\perp-y_0)$ has a
part $\Hp_{MN}(x_{||})$, given by (\ref{blurrx}), that has support
only on the brane. The remaining part $\Delta
H_{MN}(x_{||},x_\perp-y_0)$ has no support on the brane, {\it i.e.},
$\la\Delta H_{MN}\ra =0$, and is given by the decomposition,
\be
H_{MN}=\Delta H_{MN} + \frac{P(x_\perp-y_0)}{Z}\,\H_{MN}\,,
\ee  
where, $Z=\int d^nx_\perp\,P^2=(2\pi)^n\int d^nq\,\wt P^2 $. In terms
of momentum space variables,  
\be
\wt H_{MN}(k,q)=\Delta \wt H_{MN}(k,q)+\frac{\wt P(q)}{Z}\,\tH_{MN}(k)\,.
\ee  
To determine the effective action on the brane, we eliminate, first,
$\Delta H_{MN}$ by using its equation of motion and then solve for
$\H_{ij}$ and $\H_{\mu j}$ using the corresponding equations. This
leaves us with an effective action for $\H_{\mu\nu}$ or, equivalently,
an action for the $4\,$-dimensional brane variables $h^\perp_{\mu\nu},
s$, and $F^\mu$.

To integrate out $\Delta H_{MN}$, start with the bulk action
(\ref{S-bulk}), with $H_{MN}$ given by the above decomposition, and a   
Lagrange multiplier term implementing the condition 
$\la\Delta H\ra_{MN}=0$,
\be
S_{EH}^{bulk}+\int d^d x (P\Delta H_{MN})\, L^{MN}(x_{||}).
\ee
To insure gauge invariance, $L^{MN}$ must satisfy $\p_M(PL^{MN})=0$.  
The momentum space equation of motion for the variation $\delta
H_{MN}= \delta(\Delta H_{MN})$ (with no brane support) is, 
\be
\frac{A}{2}\left[p^2 \wt H^{\perp MN} +\frac{d-2}{d}(p^Mp^N-\eta^{MN}p^2)
\,\wt S \right] -\wt L^{MN}(k)\wt P(q) =0\,.
\ee  
This can be easily solved for $\wt S$ and $\wt H^{\perp MN}$ in terms
of $\wt L=\wt L^M_M$ and $\wt L^{MN}$. Blurring the solutions using
(\ref{blurr}) then determines $\wt L$ and $\wt L^{MN}$ in terms of
$\la\wt S\ra$ and $\tHp^{MN}$. Substituting back in the solutions for
$\wt S$ and $\wt H^{\perp MN}$ gives (\ref{SbrakS}) and ,
\be
\wt H^{\perp MN}=\frac{\wt P}{k^2+q^2}\,\frac{1}{\la\wt G\ra}
\left[-\tHp^{MN}+\frac{d-2}{d}\left(\frac{\la\wt g\ra^{MN}}{\la\wt
    G\ra}+\frac{p^M p^N}{k^2+q^2}\right)\la\wt S\ra\right]\,,
\ee
where we have introduced (as a generalization of (\ref{gbrak}))
\be
\la\wt g\ra^{MN}=(2\pi)^n\int d^nq\, \frac{p^Mp^N\, \wt P^2(q)}
{(k^2+q^2)^2}\,.
\ee
Substituting this back in $S_{EH}^{bulk}$ gives a $4$-dimensional
action for $\tHp^{MN}$ and $\la\wt S\ra$ (the contribution of the
Lagrange multiplier term vanishes),
\bea
S_{EH}^{bulk, eff}&=&(2\pi)^4\frac{A}{4}\int 
d^4k\frac{1}{\la\wt G\ra}\Big[\tHp^{MN}\tHp^*_{MN}-\frac{(d-2)(d-1)}
{d^2}\la\wt S\ra \la\wt S\ra^*  \nn\\
&&\hspace{6cm}-\,\frac{d-2}{d}\frac{\la\wt g\ra_{MN}}{\la\wt G\ra}
\tHp^{MN}\la\wt S\ra^*\Big]\,.
\label{bulk-eff}
\eea
From this we can further eliminate $\tHp_{ij}$ and $\tHp_{i\mu}$ by
expressing them in terms of $\la\wt S\ra$ using the solutions of the
$(i,j)$ and $(\mu, j)$ components of (\ref{eom}) along with
(\ref{SbrakS}). Then, we express $\tHp_{\mu\nu}$ and $\la\wt S\ra$
in terms of $\wt h^\perp_{\mu\nu}$, $\wt s$ and $\wt F_\mu$ using the
surface equation (\ref{Hh}) and its consequences. Note that the
equations used so far also determine $\wt F_\mu$ in terms of $\wt s$
(\ref{Fmu}), but leave $\wt s$ and $\wt h^\perp$ undetermined.
Finally, adding $S_{EH}^{brane}[h]$ leads to the $4\,$-dimensional
effective action,     
\be
S_{eff}[h]=(2\pi)^4\frac{B}{4}\int d^4k \left[\Big(-k^2 +\frac{A}{B}
\la\wt G\ra^{-1}\Big)\wt h^{\perp\mu\nu}\wt h^{\perp *}_{\mu\nu}+
\frac{3}{8}\Big(k^2 + \frac{A}{2B}\frac{d-2}{d-5} \la\wt G\ra^{-1}
\Big)\wt s\,\wt s^* \right] \,.\nn
\label{Seffh}
\ee
This gives the correct equation of motion with the correct solutions
(\ref{s}),(\ref{hp}). However, it is not yet the final form of the
effective action, as explained below.   

The above action is the analogue of the Fierz-Pauli action in the form
(\ref{SFPnoF}) which, as argued there, is not its most natural ({\it
  i.e.,} most local) form.  In fact, the $h^\perp$-terms in the two
actions can be mapped by replacing the FP mass parameter $a$ as,
\be 
a\rightarrow \wt a(k)=-\frac{A}{B}\la\wt G\ra(k)\,.
\label{wta}
\ee
Naively, the $s$-terms too can be mapped by replacing the FP $b$
parameter by $\wt a(k)/(d-6)$. But that is not the correct
comparison since it does not map the $\bF_\mu$ solution of FP theory 
(\ref{F-FPsoln}) to the $F_\mu$ solution of BIG (\ref{Fmu}) (simply
because the quantity $U$ does not appear in the map). In spite of this,
the similarity between the BIG action (\ref{Seffh}) and the FP action
in the form (\ref{SFPnoF}), teaches us the following: 

Recall that in the FP case, (\ref{SFPnoF}) was a singular limit of a
$1$-parameter family of actions (\ref{SFPw}), a generic member of
which contained the FP St\"uckelberg fields $\bF_\mu$. The $\bw_1=0$
member of this family was the original FP action (\ref{S-FP}) which
was local when written in terms of $h_{\mu\nu}$ and $\bar
f_\mu$. However, the specific $\bw_1$ choice that led to
(\ref{SFPnoF}) decoupled $\bF_\mu$, so the corresponding solution
could not be reproduced. Also this action had $1/k^2$-type
non-localities when expressed in terms of $h_{\mu\nu}$.

To compare, this also is the structure of the BIG effective action
(\ref{Seffh}).  It too does not contain the worldvolume fields $F_\mu$
and, when expressed in terms of $h_{\mu\nu}$, will contains
$1/k^2$-type non-localities arising from the projection operators
(\ref{PO})
\footnote{Of course, the action (\ref{Seffh}) also contains
  non-localities through the $k$-dependence of $k^2\la\wt G\ra$. But
  these arise as a result of integrating out bulk modes and are not
  related to St\"uckelberg fields.}. We take this as an indication 
that (\ref {Seffh}), too, is the singular limit of a $1$-parameter
family of actions that generically contain the $F_\mu$, beside the
metric. 

The fact that the replacements for $a$ and $b$ described above cannot
map the $\bF_\mu$ solution to the $F_\mu$ solution means that the
$1$-parameter family to which the BIG effective action (\ref{Seffh})
belongs cannot be obtained from the corresponding construct
(\ref{SFPw}) in FP theory. In this sense, the two $1$-parameter
families are not equivalent. The explicit construction of the BIG
$1$-parameter family is performed below and is then compared to the FP
theory for a different choice of $b$.

\subsection{A $1$-parameter family of $4$-d BIG effective actions}

To reinstate the $F_\mu$ into the BIG effective action, let's get back
to equation (\ref{bulk-eff}). We again eliminate the bulk tensors in
favour of $h^\perp$, $s$ and $F_\mu$, but now retain $F_\mu$. Since
the relation between $s$ and $F_\mu$ is already implicit in the
equations employed, the outcome is not unique and depends on the steps
followed. Depending on the details, the resulting action may not even
lead to the correct equations of motion for $h_{\mu\nu}$ and $F_\mu$.
But this can be corrected. Indeed, a candidate action that one
directly obtains in this way is {\footnote{We have rewritten $k^\mu
    k^\nu\tHp_{\mu\nu}\la\wt S\ra$ using the surface equation 
    (\ref{Hh}). Instead, writing $k^\mu k^\nu\tHp_{\mu\nu}=\la q_iq_j
    \wt H^{\perp ij}\ra$ and using the second equation in
    (\ref{trHqqH}) would have given a different result.}},  
\bea
S_{EH}^{bulk,eff}&=&(2\pi)^4\frac{A}{4}
\int d^4k\Big[\frac{1}{\la\wt G\ra}\wt h^{\perp\mu\nu}\wt h^{\perp
    *}_{\mu\nu} +\frac{U_1}{\la\wt G\ra}\wt s\wt s^*
  \nn\\ &&+\frac{2}{\la\wt G\ra}\left( k^2\wt F^2+(k\cdot \wt
  F)^2\right) +i\,\frac{U_2}{\la\wt G\ra}\left(k\cdot\wt F^*\wt
  s-k\cdot\wt F\wt s^*\right) \Big] \,,
\eea 
with, 
\bea
U_1&=&\frac{1}{16}\,\frac{(d-2)^2}{(d-1)^2(d-5)^2}\left[2(d-4)(2d-5)
-9\,\frac{d-1}{d-2}-3\,(d-7)\frac{\la\wt g\ra}{\la\wt G\ra}\right]\,,
\nn\\ U_2&=&\frac{1}{4}\,\frac{(d-2)}{(d-1)(d-5)}\left[2(d-4) -3
  -3\frac{\la\wt g\ra}{\la\wt G\ra}\right] \,.
\eea 
This does not reproduce the correct equations of motion (after adding
$S_{EH}^{brane}[h]$), indicating that the ambiguity between $s$ and
$\p\cdot F$ terms has not yet been properly resolved. To cure this,
one can use the relation between $s$ and $\p\cdot F$ in (\ref{Fmu}) to
write an appropriate part of the $\wt s^*\wt s$ term in terms of
$k\cdot\wt F$. As a shortcut, a simple inspection shows that the correct
solution $\wt F_\mu=-iU(k_\mu/4k^2)\wt s$ can be obtained if $U_2$ is,
somehow, replaced by $U$ given by (\ref{Fmu}). Hence, the trick is to
set $U_2=U+\delta U_2$ and then convert the coefficient of the $\delta
U_2$ piece to $\wt s\,\wt s^*$ form using (\ref{Fmu}). This leads to
an effective action containing $F_\mu$ that reproduces the correct
equations of motion, 
\bea
S_{eff}[h,F]&=& S_{EH}^{brane}[h]+(2\pi)^4\frac{A}{4}
\int d^4k\Big[\frac{1}{\la\wt G\ra}\wt h^{\perp\mu\nu}\wt h^{\perp
    *}_{\mu\nu} +\frac{1}{4\la\wt
    G\ra}\left(\frac{3}{4}\,\frac{d-2}{d-5}+U^2\right) \wt s\wt s^*
  \nn\\ &&\qquad+\frac{2}{\la\wt G\ra}\left( k^2\wt F^2+(k\cdot \wt
  F)^2\right) +i\,\frac{U}{\la\wt G\ra}\left(k\cdot\wt F^*\wt
  s-k\cdot\wt F\wt s^*\right) \Big]\,.  
\eea 
But, as experience with FP theory shows, this action may not be unique
in this respect. Consider converting a part of the $s s^*$ term into
$(\p F)^2$ and $\p F s$ terms, thereby, shifting the corresponding 
coefficients by $\omega_1$, $\omega_2$ and $\omega_3$ to (for
convenience we use the notation (\ref{wta})), 
\bea
&S_{eff}[h,F] = -(2\pi)^4\frac{B}{4}
\int d^4k\Big[(k^2+\wt a)\wt h^{\perp\mu\nu}\wt h^{\perp *}_{\mu\nu}-
\left(\frac{3}{8}k^2 -
\wt a(\frac{3}{16}\,\frac{d-2}{d-5}+\frac{U^2}{4}+\,w_1)\right)
\wt s\wt s^* & \nn\\[.1cm] 
&\quad +i\,(U+w_2)\wt a\left(k\cdot\wt F^*\wt s-k\cdot\wt F\wt
s^*\right)+\wt a (k_\mu\wt F_\nu-k_\nu\wt F_\mu)^2 + \wt a
(4+w_3)(k\cdot \wt F)^2 \Big]\,.&
\label{SeffhFw}
\eea 
Clearly, the modification is constrained by 
$w_1 \wt s\wt s^*+i\,w_2(k\cdot\wt F^*\wt s-k\cdot\wt F\wt s^*) +
w_3(k\cdot \wt F)^2=0$. Then the equations of motion reproduce the
correct solutions provided,   
\be 
w_3=\frac{4}{U}\, w_2  \,,\qquad  w_2=\frac{4}{U}\,w_1\,.
\ee
Hence, we have a $1$-parameter family of correct BIG effective
actions.   

The $F_\mu$ equation of motion, 
$$
2k^2\wt F_\mu +(2+w_3)\,k_\mu (k\cdot\wt F)+i(U+w_2) k_\mu\wt s
=0\,,
$$
gives the correct $F_\mu$ solution (\ref{Fmu}), except for $w_2=-U$
(implying $w_3=-4$) which leads again to the action (\ref{Seffh}). For
this special value, $\p^\mu(\p_\mu F_\nu-\p_\nu F_\mu)=0$ and $F_\mu$
is pure gauge, but otherwise undetermined. Hence, as expected,
(\ref{Seffh}) arises as the singular limit of the family
(\ref{SeffhFw}). The the complete effective action, capable of
determining all worldvolume fields $h^\perp, s$ and $F$, should be a
non-singular element of this family.

The above discussion also highlights a structural similarity between
the BIG and FP actions in the forms (\ref{SeffhFw}) and (\ref{SFPw})
which are inequivalent otherwise: The specific choices of the
parameters $w_i$ and $\bw_i$ that decouple $F_\mu$ and $\bF_\mu$,
respectively, from gravity, also convert the $F$'s into free Maxwell
fields. This, in particular, decouples the longitudinal mode of the
vector field which is ghost-like, thereby, transferring the ghost
entirely to the gravity sector. In this respect, the ghosts in BIG
(for $n>1$) and in FP theory (for $a+b\neq 0$) have similar origins,
both being related to the survival of the gauge dependent components
of the brane metric (contained in $F_\mu$).

\subsection{Spontaneous breaking of $4$-d gauge invariance}

The discussion in the previous section shows that the correct BIG
effective action is a non-singular member of the family
(\ref{SeffhFw}). The presence of $F_\mu$ in this action has the
outcome that gauge dependent components of the metric, $a_\mu$ and
$\phi$ (\ref{h-giv}), do not drop out of the action and gauge
invariance is maintained only through the compensating transformations
of $f_\mu$. Thus, the $4$-dimensional gauge invariance of the theory,
{\it i.e.}, symmetry under $\delta
h_{\mu\nu}=-2\p_{(\mu}\lambda_{\nu)}$, which is manifest in the
starting BIG action, is broken spontaneously in the effective
$4$-dimensional theory, as signaled by the need for compensating
transformations. 

This effect can be traced to the fact that the process of integrating
out non-brane components of graviton, by partially solving the
equations of motion, requires eliminating gauge degrees of freedom
through either explicit gauge fixing, or the use of gauge invariant
variables.  In our approach, this is technically related to the
necessity of using (\ref{Hh}) in relating gauge invariant bulk and
brane variables. In other approaches, the same issue will arise though
it may be obscured by an explicit gauge fixing.

To make the spontaneously broken nature of the symmetry very explicit,
we can write the effective action in terms of the gauge non-invariant
variables $h_{\mu\nu}$ and $f_\mu$. A familiar form is obtained by
matching some terms of the BIG effective action (\ref{SeffhFw}) (by
choosing an appropriate $w_1$) to the starting form of the FP action
(\ref{SFP}). For example, we can map the coefficients of the $s^2$
and $(\p F)s$ terms to FP theory by choosing  
$$
w_1=-\frac{3}{16}\frac{d-2}{d-5}\,(\frac{U}{U+2})-\frac{U^2}{4}\,.
$$
Then, in terms of 
$$
\wt a(k)=-\frac{A}{B}\la\wt G\ra(k)\,,\qquad
\wt b(k)=\frac{\wt a}{4}\left(\frac{3}{2}\,\frac{d-2}{d-5}\,
\frac{1}{U+2} \,-\, 1\right)\,,
$$
the BIG effective action takes a form very similar to the FP
action, differing from it only in the $(\p F)^2$ term. Finally, in
terms $h_{\mu\nu}$ and $f_\mu$, it becomes,
\bea
S_{eff}[h,f]&=& S_{EH}[h] -\frac{B}{4} \int d^4k \Bigg[
\wt a(k)\Big|\wt h_{\mu\nu}+2i\, k_{(\mu}\wt f_{\nu)}\Big|^2
+\wt b(k)\Big|\wt h^\mu_\mu+2i\, k_\mu\wt f^\mu\Big|^2\,
\nn\\[.2cm]
&&\hspace{3cm}
 -2\wt b(k)\,\frac{2U+1}{U}\,\Big|k\cdot\wt f+\frac{1}{6}\wt h
-\frac{2}{3}\frac{k^\mu k^\nu}{k^2} h_{\mu\nu}\Big|^2\,\Bigg] \,.
\label{SBIG-hf}
\eea
The first line is an exact map from the FP theory. The second line
shows the structural difference between massive FP gravity and the BIG 
effective action for $n>1$.  

Finally, one may comment on the similarity of brane induced gravity
with the gravitational Higgs mechanism of (\cite{'tHooft,Kakushadze}). In
these works, in order to break general covariance, one introduces $4$
scalars fields $\phi^\mu(\sigma)=\sigma^\mu+\delta\phi^\mu$, with a
Minkowski metric on the field space ($\sigma^\mu$ denote $4$
dimensional spacetime coordinates). These fields are very similar to
our brane embedding coordinates, $x^\mu(\sigma)=\sigma^\mu+f^\mu$
which naturally come with a Minkowski metric (the transverse
fluctuations $y^i$ are not relevant). In this sense, {\it BIG provides
  a natural realization of Higgs mechanism in gravity}.

Note that the presence of extra dimensions is crucial for this
similarity to hold. The reason is that, as discussed earlier, the
$f^\mu$ are non-trivial only for $n>0$. It is interesting to explore
the connection between these two theories further.

\section{Conclusions} 

Our results are summarized and discussed in detail in section 1. Here
we summarize the main conclusions. The existing classical solutions
for bulk gravity sourced by a $3$-brane tension are not adequate for
studying metric fluctuations on the brane. Hence we use the flat
background approximation but argue that the results also shed light on
the curved background analysis. In the flat approximation, brane
induced gravity filters out a cosmological constant using the mass of
the unhealthy scalar mode. This mass is partly an artifact of the flat
approximation and there is no real tachyon instability (in the sense
of an unbounded Hamiltonian) associated with it.

The presence of ghost in BIG is related to the phenomenologically
interesting values of parameters in the theory, corresponding to a
very low graviton mass and effectively $4$-dimensional gravity on the
brane (for example, there is no problem with the natural parameter
values in the string theory D-brane setups). For these parameter
values avoiding the ghost is difficult even in a curved background
sourced by $\Lambda$ since our analysis can already probe such
backgrounds for small values of $\Lambda$ and the problem is very
similar to ghosts in ordinary massive gravity. If this turns out to be
the case, then BIG cannot be regarded as a viable solution to the
cosmological constant problem.  

The only healthy exception would be a background that absorbs the
cosmological constant by curving directions transverse to the brane
while leading to {\it massless gravity} on the brane
($\Sigma_{curved}(0)=0$), if that is possible at all. This would
avoid the ghost issue and lead to a theory with filtered out
$\Lambda$ and modified massless gravity on the brane. The
modifications to gravity can be computed using the methods of section
$3.3$ and will have observable consequences. However, verifying this
possibility requires constructing brane classical solutions that
satisfy the equations of motion with the correct source terms. As
indicated earlier, finding such solutions is a generic problem in
brane physics. At the moment, we cannot say more on this possibility
(however, see see footnote $3$).

Another conclusion is that, using gauge invariant variables, it is
possible to keep track of symmetries and explicitly show that, in the
effective $4$-dimensional gravity theory, gauge dependent components
of the metric do not decouple. This indicates a spontaneous breaking
of $4$-dimensional general covariance as a consequence of the
bulk-brane setup. This setup generates the extra gauge invariant modes
that contain the St\"uckelberg fields associated with the spontaneous
breakdown of symmetry. It provides a realization of the gravitational
Higgs mechanism.

\acknowledgments

We would like to thank I. Antoniadis, M. Baumgartl,
F. Berkhahn, C. Burgess, M. Berg, C. Deffayet, J. Enander, C. de Rahm,
G. Gabadadze, M. R. Garousi, C. Germani, A. Ghodsi, H.T. Hansson,
K. Hinterbichler, J. Khoury, E. Kiritsis, P. Moyasseri, S. Ramgoolam,
R. Rosen, B. Sundborg, A. Tolley, for useful discussions and comments.

\appendix

\section{Projection operators for gauge invariant variables}

For a symmetric tensor $H_{MN}$, its transverse part $H'_{MN}$
satisfying $\p^MH'_{MN}=0$ is given by 
$$
H'_{MN}=H_{MN}-\Box^{-1}_d(\p_M\p^L H_{LN}+\p_N\p^L H_{ML}
-\Box^{-1}_d\p_M\p_N \p^L\p^K H_{LK})
$$
Then, the components in the decomposition (\ref{H-giv}) are given
by (with $H=H^L_L$),
\be
\begin{array}{ll} 
\ds H^\perp_{MN}=H'_{MN}-\frac{1}{d-1}(\eta_{MN}-\frac{\p_M\p_N}{\Box_d})
H'\,,\quad    &
\ds S=\frac{d}{d-1}(H-\frac{\p^L\p^K H_{LK}}{\Box_d})\,,
\\[.3cm]
\ds A_N=\frac{1}{\Box_d}(\p^L H_{LN}-\p_N\frac{\p^L\p^K H_{LK}}{\Box_d})
\,,\quad  &  
\ds \Phi=\frac{d}{d-1}(\frac{\p^L\p^K H_{LK}}{\Box_d\Box_d})-
\frac{1}{d-1}\frac{H}{\Box_d}
\end{array}
\label{PO}
\ee
The gauge transformation $\delta H_{MN}=-2\p_{(M}\xi_{N)}$ results
from the variation of $A_M+\frac{1}{2}\p_M\Phi$. $H^\perp_{MN}$ and
$S$ are invariant except under a restricted class of transformation
with $\Box_d\xi^M=0$. These become relevant when $\Box_d H_{MN}=0$
which is not the case in brane induced gravity.

The energy-momentum tensor $T_{MN}$ can be decomposed in a similar
way,
$$
T_{MN}=T^\perp_{MN}+\p_M T_N+\p_N T_M+\p_M\p_NT_\phi+\frac{1}{d} 
\eta_{MN} T_s
$$
The components in this expansion are given by expressions analogous to
those for $H_{MN}$ in terms of $T_{MN}$ and a $T_{MN}'$.
 
Generically, $\Box_d T_{MN}\neq 0$. Then $\p^M T_{MN}=0$ implies
$T'_{MN}=T_{MN}$ and other simplifications. However, if $\Box_d
T_{MN}=0$, then $\Box^{-1}_d\p^L\p^K T_{LK}\neq 0$ and one should not
use the simplified equations. An example is the cosmological constant
source $T^\Lambda_{MN}=\eta_{MN}\Lambda$. Then, the simplified
equations do not hold, while using the complete projection operators
one gets, $T^{\Lambda\perp}_{MN}=T^{\Lambda}_{M}=
T^{\Lambda}_{\phi}=0$ and $T^\Lambda_s=d\Lambda$.

\end{document}